%% file: paper.tex
\documentclass[prr]{revtex4-2}
\usepackage{geometry}
\usepackage{subcaption}
\usepackage{graphicx}
\usepackage{url}
\usepackage{amsmath}

\usepackage{booktabs}
\usepackage{tikz}
\usepackage{quiver}

\usetikzlibrary{matrix}
\usetikzlibrary{quantikz}
\usepackage{hyperref}
\usepackage{dsfont}
\usepackage{amssymb}
\usepackage{amsthm}
\usepackage{bm}
\newtheorem{theorem}{Theorem}

\newtheorem*{proposition*}{Proposition}

\newtheorem{definition}{Definition}

\begin{document}

\title{A Mereological Approach to Higher-Order Structure in Complex Systems:\\
from Macro to Micro with Möbius}
\author{Abel Jansma}
\email{a.a.a.jansma@uva.nl}
\affiliation{Max Planck Institute for Mathematics in the Sciences, Germany}
\affiliation{School of Informatics, University of Edinburgh, United Kingdom}
\affiliation{Dutch Institute for Emergent Phenomena, University of Amsterdam, The Netherlands}

\date{\today}

\begin{abstract}
\noindent 
Relating macroscopic observables to microscopic interactions is a central challenge in the study of complex systems. While current approaches often focus on pairwise interactions, a complete understanding requires going beyond these to capture the full range of possible interactions. We present a unified mathematical formalism, based on the Möbius inversion theorem, that reveals how different decompositions of a system into parts lead to different, but equally valid, microscopic theories. By providing an exact bridge between microscopic and macroscopic descriptions, this framework demonstrates that many existing notions of interaction, from epistasis in genetics and many-body couplings in physics, to synergy in game theory and artificial intelligence, naturally and uniquely arise from particular choices of system decomposition, or \textit{mereology}. By revealing the common mathematical structure underlying seemingly disparate phenomena, our work highlights how the choice of decomposition fundamentally determines the nature of the resulting interactions. We discuss how this unifying perspective can facilitate the transfer of insights across domains, guide the selection of appropriate system decompositions, and enable the search for new notions of interaction. To illustrate the latter in practice, we decompose the Kullback-Leibler divergence, and show that our method correctly identifies which variables are responsible for the divergence. In addition, we use Rota's Galois connection theorem to describe coarse-grainings of mereologies, and efficiently derive the renormalised couplings of a 1D Ising model. Our results suggest that the Möbius inversion theorem provides a powerful and practical lens for understanding the emergence of complex behaviour from the interplay of microscopic parts, with applications across a wide range of disciplines.

\end{abstract}

\maketitle

\tableofcontents

\section{Introduction}

Much of the study of complex systems is focused on characterizing general principles of complexity. Famously, certain network or graph theoretical quantities have proven useful to describe the structure of complex systems in fields ranging from physics to biology and the social sciences (see, e.g. \cite{essam1971graph,watts1998collective,rosen1963some,mason2007graph,pavlopoulos2011using,borgatti2009network,grandjean2016social}). In addition, concepts from a specific discipline, like phase transitions in physics or evolution in biology, have led to both qualitative and quantitative understanding of complex systems in other fields \cite{dennett1995darwin,zurek2009quantum,hodgson2005generalizing,scheffer2009early,sole2011phase}. This paper aims to introduce a new general principle of complex systems: the choice of how we decompose a system into parts can \textit{uniquely} determine the interactions among its parts. While it may seem obvious that parts must be defined before we can study how they interact, we show that this simple intuition can be made precise by representing the relationship between the parts and the whole system as a partial order. The Möbius inversion theorem then states that sums over this partial order, representing macroscopic or aggregate quantities $Q$, can be inverted through convolution with the Möbius function $\mu$ to derive the microscopic contributions $q$ \cite{Rota1964} (see Theorem \ref{thm:MIT} for the full statement): 
\begin{align}
Q(x) = \sum_{y\leq x} q(y) \iff q(x) = \sum_{y\leq x} \mu(x, y) Q(y)
\end{align}
Here, $x$ and $y$ represent elements of the partial order, corresponding to different `parts' of the system. We show that the choice of partial order (which we call a mereology, in line with the branch of philosophy and mathematical logic that studies parts and wholes) alone is sufficient to derive a complete microscopic theory of the system---whether that involves simple pairwise interactions or more complex higher-order effects. In numerous examples, we will demonstrate the effectiveness of this approach by choosing a natural mereology, and showing that this reproduces famous examples of higher-order quantities across scientific disciplines (as well as new ones). This framework thus offers a better understanding of existing theories, a precise mathematical definition for `higher-order' interactions, and a systematic way to derive new microscopic theories.

\subsection{Higher-order interactions}
It is hard to imagine a scientific question that cannot be reduced to the problem of quantifying interactions and their effects. However, interactions are commonly only studied among pairs of variables. Interactions that involve more than two variables are often collectively referred to as \textit{higher-order} interactions. Such higher-order interactions have historically been ignored for multiple reasons: their estimation typically requires more data, they are harder to interpret, and pairwise models have been surprisingly successful. In addition, pairwise interactions can be concisely represented as a graph, allowing them to be analysed with the powerful tools of graph theory. Higher-order interactions, in contrast, can only be represented as a \textit{hypergraph}, a generalisation of a graph that allows for edges to connect more than two nodes. Hypergraphs are not as well-understood as graphs, and the tools to analyse them are less developed, though significant progress is being made \cite{ghoshal2009random,lotito2022higher,leal2021forman,eidi2020ollivier}. 

While it has been shown that there are certain situations in which pairwise models are generally good approximations \cite{tkacik2006ising,merchan2016sufficiency}, higher-order interactions have proven to be of crucial importance to the rich dynamics and multistability of complex systems \cite{tanaka2011multistable,skardal2020higher,schawe2022higher}, which has inspired recent work that studied higher-order interactions in their own right \cite{battiston2020networks,battiston2021physics,rosas2022disentangling}. By now, the importance of higher-order interactions has been recognised across the sciences, motivating the attempt at a unified theory of higher-order interactions made in this paper.

In biology, for example, higher-order interactions among genetic variants and mutations have been shown to play a key role in the emergence of phenotype from genotype \cite{kuzmin2018systematic,eble2023master}. Similarly, at the level of transcription, non-additive higher-order effects control bone morphogenetic protein (BMP) signalling \cite{antebi2017combinatorial,klumpe2022context}, embryonic development of \textit{Drosophila} \cite{arnosti1996eve} and the emergence of cell type from gene expression \cite{jansma2023high,jansma2023-thesis}. On an ecological scale, certain species of lichen crucially depend on a symbiosis that involves more than two species \cite{spribille2016basidiomycete}, and higher-order interactions among species in the \textit{Drosophila} gut microbiome affect the longevity of the host \cite{gould2018microbiome}. Also in the brain, which is generally thought of as a network of pairwise connected neurons, higher-order and synergistic functional interactions among brain regions are associated with more complex and integrative cognitive processes than additive ones \cite{luppi2022synergistic}. In Section \ref{sec:bio}, each of these examples will be shown to be uniquely defined by a natural decomposition of the system.

Also in physics, higher-order interactions are ubiquitous---there is a long history of introducing non-pairwise terms into models. For example, while the Ising model on a lattice is commonly defined with only pairwise nearest-neighbour interactions, a coarse-graining or renormalisation of the lattice necessarily introduces higher-order interactions among the spins \cite{maris1978teaching}. Furthermore, spin models with varying and arbitrary-order interactions have been extensively studied in their own right, and are generally referred to as spin glasses due to their importance to the study of glassy materials \cite{jaynes1957information,gardner1985spin,kirkpatrick1987p}. In (quantum) field theory, scattering amplitudes are generally approximated by perturbative methods that sum increasing `orders' of particle interactions \cite{weinberg1995quantum,peskin2018introduction}. Both these examples will be derived as Möbius inversions in Section \ref{sec:physics}, as will the renormalisation procedure in Section \ref{sec:renormalisation}.

Finally, it should be noted that `higher-order' refers not \textit{only} to beyond-pairwise interactions. In information theory, for example, one can distinguish redundant or synergistic information among three variables. Both are defined on three variables, but synergetic terms are commonly referred to as `higher-order' \cite{ehrlich2022measure,varley2023partial} (see Section \ref{sec:infoTheory} for a precise formulation of this hierarchy). In chemistry, sums over fragments of a molecule result in `higher-order' contributions from larger submolecules, based on graph inclusion, not number of elements (see Section \ref{sec:chemistry}). Our framework is general enough to capture all these notions of higher-order structure, but precise enough to clearly distinguish between them.

\subsection{Aim and Contributions} This study demonstrates how a complex system's higher-order structure emerges uniquely from its \textit{mereological} decomposition into parts. Our approach, based on the Möbius inversion theorem, offers several key insights. First, it reveals a common mathematical structure underlying seemingly disparate notions of higher-order interactions across scientific domains. Second, it provides a general method for deriving new microscopic interactions from macroscopic observables and the transfer of insights across scientific disciplines. Third, it reveals mereology to be key in determining the nature of higher-order interactions. This perspective mathematises how the definition and estimation of interactions in a system is uniquely fixed by an assumed underlying mereology, resonating with Plato's call to ``carve Nature it at its joints" \cite{platoPhaedrus}. This unifying perspective can deepen our understanding of existing theories and guide the development of new ones.

\subsection{Related work} That the Möbius inversion theorem can be useful in the study of complex systems is itself not a novel observation. For example, Section \ref{sec:applications} will show that many instances of its use are based on the relationship between moments and cumulants of a probability distribution, which have previously been realised as Möbius inversions \cite{speed1988cumulants,rota2000combinatorics}. Furthermore, within information theory, deriving general principles of complexity based on system decompositions has been explored before in \cite{ay2011geometric}, and the authors of \cite{lang2022information,jansma2023higher} use Möbius inversion to connect different concepts from information theory. More abstractly, the relationship between mereological decompositions and emergent effects has been studied in the context of causal emergence \cite{hoel2013quantifying} and the classification of emergence in \cite{carroll2024emergence} similarly makes use of specifying a mereology. However, each of these examples only considers decompositions with the structure of a Boolean algebra, and as such essentially reduces to the set-theoretic inclusion-exclusion principle. While the partial information decomposition (Section \ref{sec:infoTheory}) does not have the structure of a Boolean algebra and has historically been referred to as a Möbius inversion, the corresponding Möbius function has only recently been identified \cite{jansma2024fast}. The approach presented in this manuscript holds for a more general class of partial orders, and it will be demonstrated that examples beyond Boolean algebras are ubiquitous.

\subsection{Structure of the Paper}
This paper is organized as follows. In Section \ref{sec:mereology}, we establish some mathematical foundations for the framework, introducing the necessary concepts from order theory and showing how the Möbius inversion theorem can be used to relate macroscopic and microscopic descriptions of complex systems. Section \ref{sec:applications} demonstrates the broad applicability of this framework by showing how it reproduces and unifies various notions of higher-order interactions across scientific disciplines, including information theory, biology, physics, chemistry, game theory, and artificial intelligence. A summary of this section is given in Table \ref{tab:summaryTable}. Having shown that the framework recapitulates existing quanties, in Section \ref{sec:kl} we present a novel application: a decomposition of the Kullback-Leibler divergence that allows for the identification of variables responsible for discrepancies between probability distributions. Section \ref{sec:renormalisation} then shows how the framework can be used to describe coarse-graining and renormalization procedures, using the 1D Ising model as an illustrative example. Finally, Section \ref{sec:discussion} discusses the implications of our results and potential future directions.

\section{Möbius Inversion as a Mereological Framework \label{sec:mereology}}
In this section, we introduce various ways to decompose a system and relate the parts to the whole. Bringing together parts to form a whole is the original meaning of algebra (\textit{al-jabr} being Arabic for reunion or rejoining of parts \cite{alKhwarizmi820}), and we indeed find that system decomposition can be fruitfully described with an algebraic technique known as a Möbius inversion. Before we present the central framework, we first define the terms and notations used in this paper.

\subsection{Preliminaries}

\begin{definition}
    Let $P$ be a set. A partial order on $P$ is a binary relation $\leq$ with the following properties for all $a, b, c \in P$:
    \begin{align}
        \text{Reflexivity: } & a \leq a \\
        \text{Transitivity: } & a \leq b \text{ and } b \leq c \implies a \leq c \\
        \text{Antisymmetry: } & a \leq b \text{ and } b \leq a \iff a=b 
    \end{align}
    Whenever $a\leq b$ we say that $a$ is \emph{less than or equal} to $b$. Two elements $a, b \in P$ are comparable when either $a\leq b$ or $b \leq a$, and incomparable otherwise. When $a\neq b$ and $a\leq b$, then we write $a < b$, and say that $a$ is \emph{less than} $b$, or $b$ is \emph{greater than} $a$.
\end{definition}
The tuple $(P, \leq)$ is referred to as a partially ordered set, or poset, but we often just write $P$ when the ordering is clear from context. Given a subset $S\subseteq P$, an element $b\in P$ is a lower bound for $S$ if $\forall s \in S: b\leq s$ (upper bounds are defined similarly). An interval $[a, b]$ on $P$ is a set $\{x: a \leq x \leq b\}$, and a poset is called locally finite if all such intervals are finite.

One can impose extra structure on the poset to form special cases. For example, a poset in which any two elements are comparable is called a totally ordered set, and a poset with a unique least (resp. largest) element is called a \textit{rooted} (resp. \textit{co-rooted}) poset. A poset $P$ in which every two elements $a, b \in P$ have a unique greatest lower bound $a \land b \in P$ and least upper bound $a \lor b$ is called a \emph{lattice}.

Given a poset $(P, \leq)$, one can study how functions on the underlying set $P$ interact with the ordering. For example, a function $f: P \to \mathbb{R}$ is called \emph{monotone} if $a \leq b$ implies $f(a) \leq f(b)$. However, there is a particular rich theory of functions on \emph{intervals} on locally finite posets, largely due to Rota \cite{Rota1964}. Functions on intervals exploit the full structure of the poset, and form an algebraic structure known as the incidence algebra, where the algebra's multiplication operation $*$ on two functions $f, g: P \times P \to \mathbb{R}$ is defined as the convolution of their values over the interval:
\begin{align}
    (f*g)(a, b) = \sum_{x:~ a\leq x \leq b} f(a, x) g(x, b)
\end{align}

Of particular interest are three elements of the incidence algebra: the \emph{delta function} $\delta_P$, \emph{zeta function} $\zeta_P$ and the \emph{Möbius function} $\mu$.

\begin{definition}[Delta function]
    Let $(P, \leq)$ be a locally finite poset. Then the delta function $\delta_P: P \times P \to \mathbb{R}$ is defined as
    \begin{align}
        \delta_P(x, y) =
        \begin{cases}
            1 & \text{if } x = y\\
            0 & \text{otherwise} \label{eq:DF}
      \end{cases}       
  \end{align}
\end{definition}
\begin{definition}[Zeta function]
    Let $(P, \leq)$ be a locally finite poset. Then the zeta function $\zeta_P: P \times P \to \mathbb{R}$ is defined as
    \begin{align}
        \zeta_P(x, y) =
        \begin{cases}
            1 & \text{if } x\leq y\\
            0 & \text{otherwise} \label{eq:ZF}
      \end{cases}       
  \end{align}
\end{definition}
\begin{definition}[Möbius function]
    Let $(P, \leq)$ be a locally finite poset. Then the Möbius function $\mu_P: P \times P \to \mathbb{R}$ is defined as
    \begin{align}
        \mu_P(x, y) =
        \begin{cases}
            1 & \text{if } x=y\\
            -\sum\limits_{z: x\leq z < y}\mu_P(x, z) & \text{if } x < y  \\
            0 & \text{otherwise} \label{eq:MF}
      \end{cases}       
  \end{align}
\end{definition}
When the underlying poset is clear from context or irrelevant, we sometimes omit the subscript $P$. Note that $\delta$ corresponds to the multiplicative unit ($\delta * f = f * \delta= f$ for all $f$ in the incidence algebra), and that multiplying a function $f: P \to \mathbb{R}$ with the zeta function amounts an integral over the poset: $(f*\zeta)(x, y) = \sum_{x\leq z \leq y} f(z)$ (to properly define the multiplication, interpret $f$ as a function on $P\times P$ that is constant in its first argument). One of the most important results in the theory of incidence algebra then states that the Möbius function is the multiplicative inverse of the zeta function ($\mu * \zeta = \zeta * \mu = \delta$) and that therefore the following theorem holds:

\begin{theorem}[Möbius inversion theorem, Rota \cite{Rota1964} \label{thm:MIT}]
Let $P = (S, \leq)$ be a locally finite poset and $\tau, \eta \in S$. Let $f: P \rightarrow \mathbb{R}$ be a function on $P$, and let $\mu_P$ be the Möbius function on $P$. Then
\begin{align}
    f(\tau) &= \sum_{\eta \leq \tau} g(\eta) \quad \iff \quad g(\tau) = \sum_{\eta \leq \tau} \mu_P(\eta, \tau) f(\eta)
\end{align}
\end{theorem}

The Möbius inversion theorem states that sums over a poset can be inverted by looking up the Möbius function of the poset. This powerful result forms the basis of the framework presented in this study. In the following sections, we will show how this theorem can be applied to a wide range of systems, and how it can be used to define and estimate many well-established notions of higher-order structure in complex systems.

\subsection{Decomposing Systems from Macro to Micro}
Given a system $S$ with parts $s_i$, consider an arbitrary property $Q(S)$ of the system. A purely additive property would be one where 
\begin{align}
    Q(S) = \sum_i q(s_i) \label{eq:atomicDecomp}
\end{align}
where $q(s_i)$ is the contribution of the part $s_i$ to the property $Q(S)$. Since $Q(S)$ is built from contributions of the parts, we think of $Q$ as describing a macroscopic (or global) property, while $q$ describes microscopic (or local) quantities or interactions. For example, the height of a person with two genetic variants can be written as $H(\{g_1, g_2\}) = h(\emptyset) +  h(\{g_1\}) + h(\{g_2\})$, where $h(s_i)$ is the contribution of the genetic variants $s_i$ to a person's total height $H(\{g_1, g_2\})$. Note the difference in interpretation between $H$ and $h$ here: $H$ is the length of a person and easy to measure, whereas $h$ is the length added by a genetic variant which is generally impossible to directly observe.

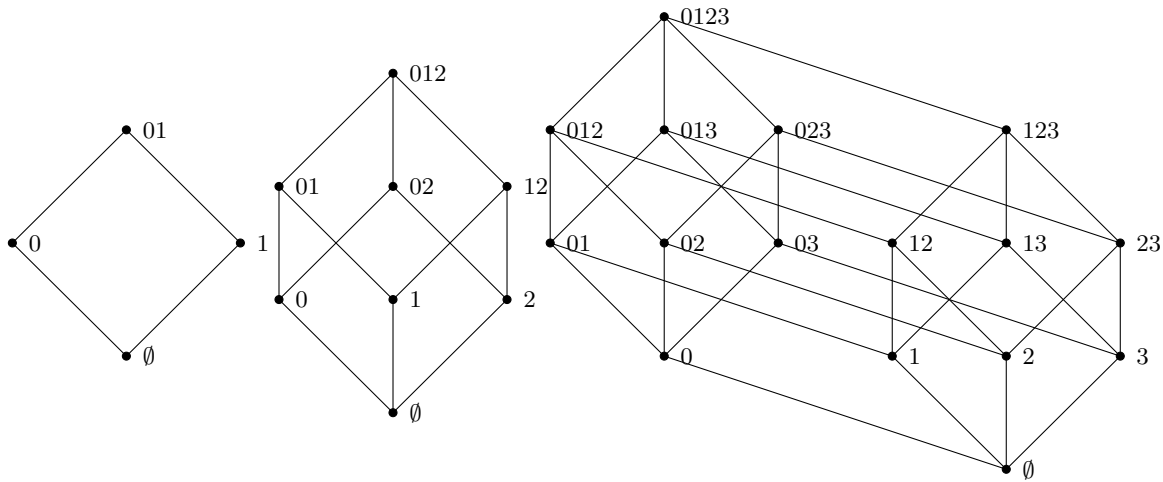
\begin{figure}
    \begin{subfigure}[t!]{0.2\textwidth}
        \centering
        \input{booleanLatticeN2.tex}
    \end{subfigure}
    \begin{subfigure}[t!]{0.2\textwidth}
        \centering
        \input{booleanLatticeN3.tex}
    \end{subfigure}
    \begin{subfigure}[t!]{0.55\textwidth}
        \centering
        \input{booleanLatticeN4.tex}
    \end{subfigure}
    \caption{The powerset of a set of $n$ variables, ordered by set inclusion, forms a lattice known as a Boolean algebra. Shown here are the transitive reductions (Hasse diagrams) of the Boolean algebras on 2 (left), 3 (middle) and 4 (right) variables. For arbitrary $n$, the Hasse diagrams of a Boolean algebra describes an $n$-cube.  \label{fig:lattice_BA}}
\end{figure}

\begin{figure}
    \begin{subfigure}[t!]{0.1\textwidth}
        \centering
        \input{partitionLatticeN2.tex}
    \end{subfigure}
    \begin{subfigure}[t!]{0.2\textwidth}
        \centering
        \input{partitionLatticeN3.tex}
    \end{subfigure}
    \begin{subfigure}[t!]{0.59\textwidth}
        \centering
        \input{partitionLatticeN4.tex}
    \end{subfigure}
    \caption{The partitions of a set of $n$ variables, ordered by refinement, form a lattice. Shown here are the transitive reductions (Hasse diagrams) of the partition lattices on 2 (left), 3 (middle) and 4 (right) variables. \label{fig:lattice_Part}}
\end{figure}
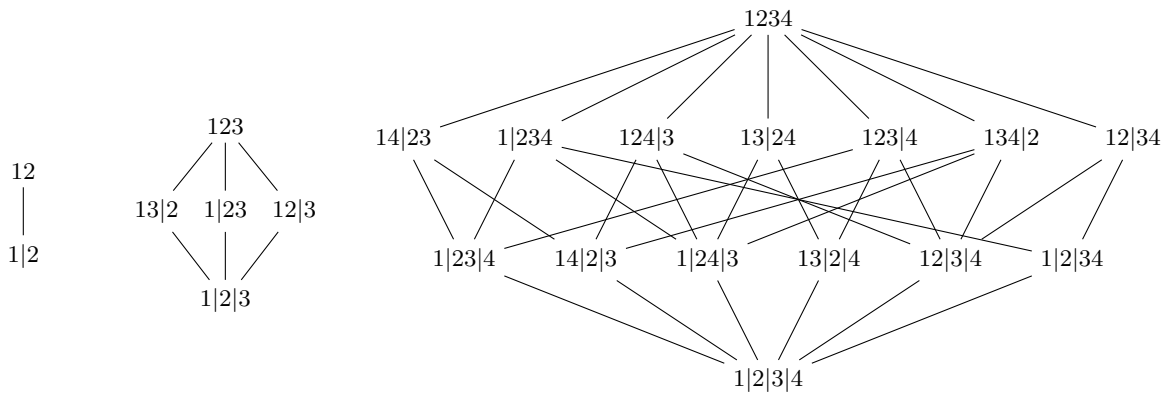

However, many interesting properties of complex systems are non-additive and emerge from complex interactions among the components. Still, it is reasonable to assume that $Q$ could be decomposed into properties of the components in a nonadditive or interacting way (as anything else would entail $Q$ being \textit{strongly} emergent and \textit{fundamentally} not derivable from a description of its parts). To allow for nonadditive effects, we expand the summand in Equation \eqref{eq:atomicDecomp} to include contributions from more complex, non-atomic parts of the system. Note that this is somewhat analogous to introducing nonlinear terms in the design matrix of a linear regression problem. This invites one to instead write 
\begin{align}
    Q(S) = \sum_{t \in \mathcal{D}(S)} q(t) \label{eq:sumOfDecomp}
\end{align}
where $\mathcal{D}(S)$ is some kind of decomposition of $S$ that relates the whole to its parts. When $S$ has a set-like structure, two natural choices for $\mathcal{D}(S)$ would be the set of all subsets (i.e. the powerset) or partitions of $S$. These two decompositions can be partially ordered by inclusion and refinement, respectively, as shown in Figures \ref{fig:lattice_BA} and \ref{fig:lattice_Part}. Choosing a particular $\mathcal{D}(S)$ amounts to a mereological claim about the system $S$ and property $Q$, so should be informed by prior knowledge of the system or the method of observation. For this reason, we refer to such system decompositions as \textit{poset mereologies}:
\begin{definition}[Poset Mereology]
    A poset mereology $\mathcal{D}(S)$ on a system $S$ is a locally finite co-rooted poset with $S$ as its unique largest element.
\end{definition}
Throughout this paper, we will refer to poset mereologies simply as mereologies. Note that in particular every topology on a finite set $S$ gives rise to a mereology on $S$ through ordering the open sets by inclusion. However, a mereology is more general in the sense that it need not be closed under unions or intersections. In fact, the full system is not required to be a set at all (it can, for example, be an ordered list, or a graph). A mereology is simply a way to collect all relations between the whole $S$ and all its relevant parts. One could drop the requirement that a mereology be locally finite, allowing for continuous mereologies, but we will not consider these in this paper.


\subsection{Inverting decompositions with Möbius functions \label{sec:inversion}}
If $\mathcal{D}(S)$ is a mereology, Equation \eqref{eq:sumOfDecomp} can be written as
\begin{align}
    Q(S) = \sum_{t \leq S} q(t) \label{eq:sumOfPoset}
\end{align}
where the sum is now over elements from the mereology. If the microscopic contributions $q(t)$ are known, then predicting the macroscopic quantity $Q(S)$ is known as the \emph{forward problem}. However, in many cases we might only be able to observe $Q(S)$, and not the individual contributions $q(t)$. To reverse-engineer $q(t)$ from $Q(S)$, Equation \eqref{eq:sumOfPoset} should be inverted to express $q(t)$ in terms of observations of $Q$ on different elements of the mereology. To invert sums like equation \eqref{eq:sumOfPoset} over posets, one can use the Möbius inversion theorem:
\begin{align}
    q(S) = \sum_{t \leq S} \mu_{\mathcal{D}(S)}(t, S) Q(t) \label{eq:qFromQ}
\end{align}
This is a powerful result: the problem of inverting the decomposition (which amounts to solving a large system of equations) has been reduced to looking up the Möbius function of the mereology. 

This statement can be summarised diagrammatically as follows. Let $\Sigma$ denote a class of systems of interest, and let $\mathcal{D}(\Sigma) = \{\mathcal{D}(S)|S \in \Sigma\}$ denote the set of $\mathcal{D}$-mereologies on systems in $\Sigma$. Let $\mathbb{R}^{\mathcal{D}(\Sigma)}$ be the set $\{f| f: \mathcal{D}(S) \to \mathbb{R}, S \in \Sigma\}$ of real-valued functions on mereologies on $\Sigma$. Let $\mathcal{F}_g:\mathcal{D}(\Sigma) \to \mathbb{R}^{\mathcal{D}(\Sigma)}$ be the function that picks out the function $g: \mathcal{D}(S) \to \mathbb{R}$ for each system $S$. The two definitions $q$ and $Q$ are called $\mathcal{D}$-compatible if the following diagram commutes:

\begin{tikzpicture}[auto]
    \centering
    \begin{scope}[xshift=-4cm]
    \node (B) at (2,0) {$\mathcal{D}(\Sigma)$};
    \node (C) at (4,0) {$\mathbb{R}^{\mathcal{D}(\Sigma)}$};
    \node (D) at (2,-2) {$\mathbb{R}^{\mathcal{D}(\Sigma)}$};
    
    \draw[->] (B) to node {$\mathcal{F}_q$} (C);
    \draw[->] (B) to node [swap] {$\mathcal{F}_Q$} (D);
    \draw[->] (C) to [bend left=35] node {$\zeta_\mathcal{D} * \_ $} (D);
    \draw[->] (D) to [bend left=25] node [swap] {$\mu_\mathcal{D} * \_ $} (C);
    \end{scope}
    
    \node at (3.5,-1) {which on elements looks like};
    
    \begin{scope}[xshift=6cm]
        \node (B) at (2,0) {$\mathcal{D}(S)$};
        \node (C) at (4,0) {$~q~~~$};
        \node (D) at (2,-2) {$~~Q~$};
        
        \draw[|->] (B) to node {$\mathcal{F}_q$} (C);
        \draw[|->] (B) to node [swap] {$\mathcal{F}_Q$} (D);
        \draw[|->] (C) to [bend left=45] node {$\zeta_\mathcal{D} * \_ $} (D);
        \draw[|->] (D) to [bend left=20] node [swap] {$\mu_\mathcal{D} * \_ $} (C);
    \end{scope}
\end{tikzpicture}

If $Q$ and $q$ are $\mathcal{D}$-compatible, then the forward problem is solved by multiplying $q$ by $\zeta_\mathcal{D}$, and the inverse problem is solved by multiplying $Q$ by $\mu_\mathcal{D}$. In fact, the Möbius inversion theorem states that for any macroscopic observable $Q$ there is a \emph{unique} notion of microscopic higher-order interaction $q$ that is $\mathcal{D}$-compatible with $Q$, and \emph{vice versa}. 

Since different systems might admit similar mereologies, the Möbius function can be precomputed and the inversion can be done in a single step. In practice, the most common mereologies on a system are the powerset and partition lattices from Figures \ref{fig:lattice_BA} and \ref{fig:lattice_Part}, and the Möbius function of these lattices is well-known. Denoting the powerset lattice by $\mathcal{P}(S)$, and the partition lattice by $\Pi(S)$, the Möbius functions are given by
\begin{align}
    \mu_{\mathcal{P}(S)}(x, y) &= 
    \begin{cases}
        (-1)^{|y|-|x|} & \text{if } x \leq y\\
        0 & \text{otherwise}
    \end{cases} \label{eq:MF_BA}\\
    \mu_{\Pi(S)}(x, S) &= (-1)^{|x|-1} (|x|-1)!
\end{align}
where $|x|$ is the cardinality of the set $x$. These expressions are well-known and have been used to invert decompositions across scientific disciplines, as will be discussed in more detail in Section \ref{sec:applications}.

In the example of genetic variants determining a person's height, let us imagine decomposing a person's height over their genotype with the powerset mereology as $H(\{g_1, g_2\}) = h(\emptyset) +  h(\{g_1\}) + h(\{g_2\}) + h(\{g_1, g_2\})$. Note that this now includes an interaction term $h(\{g_1, g_2\})$. Inserting Equation \eqref{eq:MF_BA} into \eqref{eq:qFromQ} allows us to calculate the effect of genetic variants from observations of a population's height. Omitting curly brackets for clarity, we find
\begin{align}
    h(g_1) &=   (g_1) - H(\emptyset) \\
    h(g_2) &= H(g_2) - H(\emptyset) \\
    h(g_1, g_2) &= H(g_1, g_2) - H(g_1) - H(g_2) + H(\emptyset)\\
    &= (H(g_1, g_2) - H(g_2)) - (H(g_1) - H(\emptyset))
\end{align}
This is easily interpreted: the effect of a single genetic variant is the difference between the height of a person with only that variant and the height of a person without any of the variants. The interaction among two variants $g_1$ and $g_2$ is the difference between the effect of $g_1$ in people with $g_2$, and the effect of $g_1$ in people without $g_2$. Given a population sample containing people with all combinations of these variants, both of these quantities can be directly estimated. This argument straightforwardly extends to higher-order effects: the third-order interaction among three genetic variants $g_1, g_2, g_3$ is given by
\begin{align}
    h(g_1, g_2, g_3) = H(g_1, g_2, g_3) - H(g_1, g_2) - H(g_1, g_3) - H(g_2, g_3) + H(g_1) + H(g_2) + H(g_3) - H(\emptyset)
\end{align}
In genetics, interaction terms like $h(g_1, g_2)$ are called epistatic effects and commonly defined and estimated using exactly this estimator \cite{gould2018microbiome,beerenwinkel2007epistasis} (see Section \ref{sec:bio}), though not generally linked to Möbius inversions. Note that the bottom elements of the mereology, namely the sets $\{\emptyset, \{g_1\}, \{g_2\}\}$, correspond to a reductionistic theory, i.e. a theory without higher-order interactions. This principle holds in general: reductionistic theories correspond to theories based on only the bottom part of a mereology. Higher-order interactions are thus `higher' with respect to the ordering of the mereology, which allows one to compare and order different interactions and theories.

This is further illustrated by the examples presented in Section \ref{sec:applications}, where we show in more detail how this construction can be applied to systems with a clear mereological structure, and how it reproduces many well-established notions of higher-order structure in complex systems.

\section{Möbius Inversions in Complex Systems \label{sec:applications}}

We aim to show that the framework presented in Section \ref{sec:mereology} reproduces many established notions of higher-order interactions throughout the sciences. First, the role of Möbius inversions in information theory is discussed in Section \ref{sec:infoTheory}. Within the natural sciences, multiple kinds of interactions in biology (Section \ref{sec:bio}), physics (Section \ref{sec:physics}), and chemistry (Section \ref{sec:chemistry}) are discussed. We discuss the role that Möbius inversions play in game theory and artificial intelligence in Sections \ref{sec:gameTheory} and \ref{sec:ml}, respectively. An overview of all dualities between macroscopic and microscopic quantities is given in Table \ref{tab:summaryTable} of the Discussion. The wide range of examples covered here serve mainly to illustrate the broad applicability of the framework, but it is by no means necessary to understand the details of all of them to appreciate the general idea. Some of the examples in this section are related to the relationship between statistical moments and cumulants. We chose to focus on scientific applications of the framework, but a description of various kinds of moments and cumulants in terms of mereological decompositions is included in Appendix \ref{sec:stats} for completeness. 

\subsection{Information theory \label{sec:infoTheory}}
\subsubsection{Entropy and Mutual Information}
Given a set $X$ of random variables $X_i$, the joint entropy $H(X)$ is the total amount of uncertainty about the state of $X$ before an observation, or equivalently, the amount of information gained by observing $X$ to be in a particular state $x$. It is defined as
\begin{align}
    H(X) = - \sum_{x \in \mathcal{X}} p(X=x) \log p(X=x)
\end{align}
where $\mathcal{X}$ is the set of all possible states of $X$, and $p(X=x)$ is the probability of observing state $x$. One might assume that the total information in the system can be decomposed into information contained in different parts of the system. To do this, we impose the powerset mereology on $X$. The joint entropy can then be written as
\begin{align}
    H(X) = \sum_{S \in \mathcal{P}(X)} I(S)
\end{align}
where $I(S)$ is the information contributed by the part $S$, given by a Möbius inversion over the powerset mereology on $X$:
\begin{align}
    I(X) &= \sum_{S \subseteq X} \mu_{\mathcal{P}}(S, X) H(S)\\
    &=  \sum_{S \subseteq X} (-1)^{|S|-|X|} H(S)
\end{align}
For two variables $X_1$ and $X_2$ this yields
\begin{align}
    I(X_1, X_2) &= H(X_1, X_2) - H(X_1) - H(X_2)
\end{align}
This is, up to a minus sign, exactly the definition of mutual information, which is the amount of information shared between two variables (or equivalently: the Kullback-Leibler divergence between the joint distribution and the product of the marginals):
\begin{align}
    I(X_1, X_2) = \sum_{x_1, x_2} p(x_1, x_2) \log \frac{p(x_1, x_2)}{p(x_1) p(x_2)}
\end{align}
In general, the mutual information among a set of variables, also referred to as their \textit{interaction information}, is indeed given by the Möbius inversion of the entropy of the powerset of the variables, with some authors adding a minus sign to even orders to ensure a positive sign for single-variable entropies. This is a well-established result, and has historically been explained through an analogy between Shannon information theory and set theory \cite{yeung1991new}.

In fact, the same argument holds for the pointwise information, or surprisal $h(x) = - \log p(X=x)$, which is the amount of information gained by observing a particular realisation $X=x$. A Möbius inversion on $\mathcal{P}(X)$ then leads to the definition of pointwise mutual information:
\begin{align}
    i(X_1, X_2) = \log \frac{p(x_1, x_2)}{p(x_1) p(x_2)}
\end{align}
This construction has previously been discussed in more detail in \cite{jansma2023higher}. Note that when only the smallest elements of the powerset mereology---the singleton sets---have non-zero interactions, the theory reduces to its `reductionistic' version where all variables are independent. 

The powerset decomposition of entropy and surprisal is the simplest option, by far the most common, and the basis of all Shannon information theory. However, in recent years, other decompositions have been proposed, motivated at least in part by the fact that higher-order mutual information can become negative, which has hindered its operational interpretation. One of these alternative decompositions is called the \textit{partial information decomposition}.

\subsubsection{The Partial Information Decomposition}
\begin{figure}
    \begin{minipage}{\textwidth}
        \centering
        \includegraphics[width=0.9\textwidth]{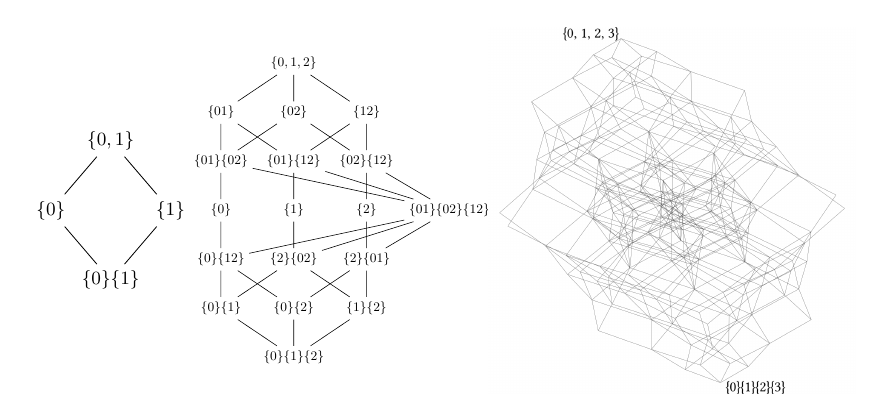}
        \caption{The redundancy terms of a powerset of $n$ variables can be partially ordered: if $A$ and $B$ are redundancy terms, then $A\leq B$ if for every $b\in B$ there is an $a \in A$ such that $a\leq b$. Shown here are the transitive reductions (Hasse diagrams) of the redundancy lattices on 2 (a), 3 (b) and 4 (c, labels not shown) variables. Note that redundancies among $n$ variables are the antichains in the powerset of $n$-variables.\label{fig:lattice_Red}}
    \end{minipage}
\end{figure}

The partial information decomposition (PID) framework, introduced by Williams and Beer \cite{williams2010nonnegative}, proposes that the information a set of predictor variables $X$ contains about a target $Y$ can be decomposed into various terms representing synergistic information (available only in the joint state of the predictor variables), unique information (exclusively contained in a single predictor variable), and redundant information (shared among multiple predictor variables). For instance, given two predictor variables $X_1$ and $X_2$, the information they carry about $Y$, denoted $I(\{X_1, X_2\}; Y)$, can be decomposed as follows:
\begin{align}
    I(\{X_1, X_2\}; Y) = \Pi(\{X_1\}; Y) + \Pi(\{X_2\}; Y) + \Pi(\{X_1\}\{X_2\}; Y) + \Pi(\{X_1, X_2\}; Y)
\end{align} 
where $\Pi(\{X_i\};Y)$ represents the unique information carried by $X_i$ about $Y$, $\Pi(\{X_1, X_2\};Y)$ denotes the synergistic information available only from the joint state of the variables, and $\Pi(\{X_1\}\{X_2\}; Y)$ represents the redundant information about $Y$ shared by $X_1$ and $X_2$. To construct information sources from an arbitrary set $S$ of predictor variables, one should consider redundancies among all possible combinations of subsets of $S$. However, note that the redundancy among a set $a$ and a set $b$ reduces to the unique information of $a$ if $a \subseteq b$. This restricts the set of information sources to combinations of predictor subsets that are mutually incomparable by the ordering $\subseteq$. Such incomparable sets are called \textit{antichains} of the poset $(\mathcal{P}(S), \subseteq)$. The antichains are then turned into the redundancy mereology $\mathcal{R}_n$ by setting $A\leq B$ if for every $b\in B$ there is an $a \in A$ such that $a\subseteq b$. The Hasse diagram of this lattice of antichains is shown in Figure \ref{fig:lattice_Red} for up to $|S|=4$. The information $I(S; Y)$ that a set of variables $S$ carries about $Y$ can then be written as a sum of `partial' information contributions $I_\partial$ over the $\mathcal{R}_n$ mereology:
\begin{align}
    I(S; Y) = \sum_{R \leq S} I_\partial(R; Y)
\end{align}
This can be inverted to get expressions of the individual contributions $I_\partial(R; Y)$ in terms of the mutual information $I(T; Y)$:
\begin{align}
    I_\partial(R; Y) = \sum_{T \leq R} \mu_{\mathcal{R}_n}(T, R) I(T; Y) \label{eq:PID_inversion}
\end{align}
Again, if only the bottom element $I_\partial(\{\{t\}\mid t \in S\}; Y)$ is non-zero, the theory reduces to its `reductionistic' version where all information is fully redundant. The inversion in Equation \eqref{eq:PID_inversion}, however, is far from trivial as $|\mathcal{R}_n|=\mathcal{D}(n)$, where $\mathcal{D}(n)$ is the $n$th Dedekind number. The series of Dedekind numbers grows so quickly that it is only known up to $D(9)$ \cite{jakel2023computation,van2023computation}. This makes naively solving the system of associated equations computationally infeasible for large $n$. Nevertheless, based on the presented framework, a closed-form expression for the Möbius function on the redundancy lattice has recently been derived and used to calculate the PID where this was not possible before \cite{jansma2024fast}.

Note, however, that even when the Möbius function can be calculated, there is an ambiguity in the definition of $I(T; Y)$ when the antichain $T$ contains more than a single set of variables. Consequently, solving the PID requires a well-defined notion of information on arbitrary antichains. A significant portion of the literature on the PID has focused on constructing such definitions.

Recently, the PID framework has been extended to accommodate multiple target variables and information dynamics in time \cite{mediano2021towards}. The associated $\Phi$ID mereology is the product of the standard redundancy mereologies associated to the individual targets. Since the Möbius function of a product of lattices is the product of Möbius functions of the lattices \cite{stanley2011enumerative}, the $\Phi$ID calculation reduces to that of the normal PID \cite{jansma2024fast}.

\subsection{Biology \label{sec:bio}}
\subsubsection{Epistasis}
In the last decade, improved genetic sequencing of genomes and transcriptomes has revealed that some traits depend on many, if not most genes. This change in perspective has been described as a transition from monogenic, to polygenic and omnigenic models \cite{boyle2017expanded}. Furthermore, the effect of a single genetic variant can depend on the presence of other genetic variants, a type of genetic interaction called epistasis. As in the example of a person's height in Section \ref{sec:mereology}, a general phenotype $F(G)$ that depends on the presence ($g_i=1$) or absence ($g_i=0$) of a set of genetic variants $g_i \in G$ can be decomposed into the sum of the effects of subsets $g_i\in\mathcal{P}(G)$ of genetic variants. A Möbius inversion on the lattice of subsets of genetic variants gives the following definitions for genetic effects and epistatic interactions $I(g)$:
\begin{align}
    F(g=\bm{1}, G\setminus g=\bm{0}) &= \sum_{s\in\mathcal{P}(g)} I(s) \iff I(g) = \sum_{s\subseteq g} (-1)^{|s| - |g|} F(s=\bm{1}, G\setminus s=\bm{0})
    \intertext{That is, the interaction among a set of genes $g$ can be expressed in terms of observed phenotypes of individuals with varying genotypes. For example, the interaction among three genes $g_1, g_2, g_3$ is given by}
    I_{g_1 g_2 g_3} = I(\{g_1, g_2, g_3\}) &= F(1, 1, 1, \bm{0}) - F(1, 1, 0, \bm{0}) - F(1, 0, 1, \bm{0}) - F(0, 1, 1, \bm{0}) \\
    \nonumber &+ F(1, 0, 0, \bm{0}) + F(0, 1, 0, \bm{0}) + F(0, 0, 1, \bm{0}) - F(0, 0, 0, \bm{0})
\end{align}
This expression is indeed commonly used to estimate epistasis. The authors of \cite{beerenwinkel2007epistasis} arrive at a similar expression based on volumes of polytopes, which has been used to describe epistasis among genetic variants \cite{eble2023master} as well as in the context of microbiomes \cite{gould2018microbiome}. 

Even though interactions are necessarily defined with respect to some outcome, this need not be a phenotype in the traditional sense of the word. One possible outcome that requires no macroscopic measurements, for example, is the population probability of a particular gene expression pattern. If we represent gene expression as a binary vector $\bm{G}$, where $G_i=1$ if gene $i$ is expressed above some threshold, then inverting the log-probability of observing transcriptome $\bm{G}$ gives the following definition of the interaction among genes $g_i$:
\begin{align}
    I(\bm{g}) &= \sum_{\bm{s}\subseteq \bm{g}} (-1)^{|\bm{s}| - |\bm{g}|} \log p(\bm{s}=\bm{1}, \bm{g}\setminus \bm{s}=\bm{0})
\end{align}
which can be directly estimated from a population sample. A method to estimate such interactions from single-cell RNA-seq data was recently developed, and used to identify novel cell states and types in various organisms \cite{jansma2023high}.

Instead of inferring the log-probability, i.e. the surprisal, one could also invert its expected value, i.e. the entropy. As explained in Section \ref{sec:infoTheory}, this yields (higher-order) mutual information, which is used by the algorithm that won the DREAM2 challenge of inferring gene regulatory networks from expression data \cite{watkinson2009inference}.

\subsubsection{Epidemiology}
The central challenge in epidemiology is to understand the influence of various factors on an individual's health outcomes. One example is estimating the effect a vaccine has on disease risk. However, disease risk after vaccination is not only determined by the vaccine itself, but also by innate factors like the individual's immune system and external factors like the presence of other pathogens, as well as interactions among these. Typically, the effect on outcome $Y$ of a factor $X$ is isolated from all other factors by randomly dividing the population into groups, and then intervening such that $X=1$ in group 1, and $X=0$ in group 2. The \textit{Average Treatment Effect} (ATE) is then defined as
\begin{align}
    ATE(Y;X) = E(Y|X=1) - E(Y|X=0)
\end{align}
However, in observational studies, where randomisation followed by intervention is not possible, the effect of $X$ generally cannot be isolated, and the outcome $Y$ depends on a potentially large set of factors $S$ on which one might impose the powerset mereology:
\begin{align}
    E(Y|S=1) = \sum_{s\in\mathcal{P}(S)} I(s) 
\end{align}
where $I(s)$ now represents the effect of the collection of factors $s$ on the outcome $Y$. The effect of a single factor $s=X$ is then seen to be given by an ATE in the absence of the other factors:
\begin{align}
    I(X) &= E(Y|X=1, S\setminus X=0) - E(Y|X=0, S\setminus X=0)
    \intertext{whereas the second- and higher-order terms correspond to higher-order epidemiological interactions among the factors \cite{beentjes2020higher}. For example, the interaction among two factors $X_1, X_2$ is given by}
    I(X_1, X_2) &= E(Y|X_1=1, X_2=1, S\setminus \{X_1,X_2\}=0) - E(Y|X_1=1, X_2=0, S\setminus \{X_1,X_2\}=0) \\
    \nonumber &- E(Y|X_1=0, X_2=1, S\setminus \{X_1,X_2\}=0) + E(Y|X_1=0, X_2=0, S\setminus \{X_1,X_2\}=0)\\
    &= ATE(Y;X_1 | X_2=1, S\setminus \{X_1,X_2\}=0) -ATE(Y;X_1 | X_2=0, S\setminus \{X_1,X_2\}=0)
\end{align}
In theory, this allows one to estimate drug or treatment interactions from population statistics, though without further knowledge of the conditional dependencies or the causal graph the interactions are not guaranteed to reflect the true `causal' relationships.  

\subsubsection{Neuroscience}
The history of neuroscience can accordingly be described by evolving views on what the appropriate mereological decomposition of the brain is: from neurons, minicolumns, neural circuits, and hierarchical structures, to the currently popular approach that focuses on different spatial and functional regions \cite{bechtel2002decomposing,sporns2005human}. To describe the correlations and higher-order relationships between parts of the brain, both the classical approach to higher-order information theory and the PID are commonly used, reflecting concurrent use of both the powerset and redundancy mereology \cite{erramuzpe2015identification,ince2017statistical,wibral2017quantifying,sherrill2021partial,newman2022revealing}

\subsection{Physics \label{sec:physics}}

\subsubsection{Equilibrium dynamics \label{sec:equilDyn}} Statistical physics is largely focused on relating the large-scale behaviour of a system to the microscopic interactions. Commonly, this is done through an energy function $E: \mathcal{S} \to \mathbb{R}$ that maps a state $s \in \mathcal{S}$ of the system to its energy. Writing down a form of $E$ amounts to choosing a decomposition of the system into parts. For example, the Ising model assumes that the energy of a system $S=\{s_1, \ldots, s_n\}$, $s_i \in \{0, 1\}$, decomposes into singleton and pairwise contributions and is given by
\begin{align}
    E(S=s) = -\sum_{i, j=1}^n J_{ij} s_i s_j - \sum_{i=1}^n h_i s_i
    \intertext{or, for a subsystem $\hat{S} \subseteq S$}
    E(\hat{S}=\bm{1}, S\setminus \hat{S}=\bm{0}) = -\sum_{i, j:~ s_i, s_j \in \hat{S}} J_{ij} - \sum_{i:~ s_i \in \hat{S}}^n h_i
\end{align}
where $h_i$ is the external field acting on spin $i$ and $J_{ij}$ is the interaction strength between spins $i$ and $j$, typically only non-zero for nearest neighbours on a lattice. Limiting the interactions to linear and pairwise quantities only is an assumption imposed by the Ising model. A more general form of the energy function would be
\begin{align}
    E(S=s) = -\sum_{t \in \mathcal{P}(S)} J_t \prod_{i: S_i \in t} s_i
\end{align}
where $J_t$ is the interaction strength of the part $t$. If one is given the parameters $J_t$, the \textit{forward Ising problem} is calculating the behaviour of the system $S$ under the influence of the energy function $E$. The \textit{inverse Ising problem} is inferring the parameters $J_t$ from observations of $S$. The inverse problem is in general intractable, as direct observations of the energy are not possible. However, at equilibrium, the probability of observing a state $s$ is given by the Boltzmann distribution
\begin{align}
    p(s) = \frac{1}{Z} e^{-E(s)}
\end{align}
where $Z=\sum_{s} e^{-E(s)}$ is the partition function. This means that one can observe the energy (up to an unimportant global shift) of a state indirectly as $-\log p(s)$ by simply estimating the probability of a state from a collection of samples. From this, a Möbius inversion quickly yields exact values for the parameters $J_t$ \cite{jansma2023higher}:
\begin{align}
    E(S=\bm{1}) &= \sum_{t \in \mathcal{P}(S)} J_t \\
    \iff J_S &= \sum_{t \leq S} \mu_P(t, S) \log p(t=\bm{1}, S\setminus t = \bm{0})\\
    &= \sum_{t \leq S} (-1)^{|t|-|S|}\log p(t=\bm{1}, S\setminus t = \bm{0})
\end{align}
For example, the inverse Ising problem for pairwise nearest-neighbour interactions at equilibrium is solved by
\begin{align}
    J_{ij} &= \log \frac{p(s_i=1, s_j=1, s_{-(i, j)}=\bm{0})p(s_i=0, s_j=0, s_{-(i, j)}=\bm{0})}{p(s_i=1, s_j=0, s_{-(i, j)}=\bm{0})p(s_i=0, s_j=1, s_{-(i, j)}=\bm{0})}\\
    \intertext{where $s_{-(i, j)}$ denotes all spins except $i$ and $j$. Simplifying notation further by writing $p_{abc} = p(s_i=a, s_j=b, s_k=c, s_{-(i, j, k)}=\bm{0})$, the 3-point coupling is given by}
    J_{ijk} &= \log \frac{p_{111}p_{100}p_{010}p_{001}}{p_{000}p_{011}p_{101}p_{110}} \label{eq:ising_3point}
\end{align}
This solution to the inverse problem was already noted in \cite{beentjes2020higher}, but the argument presented here shows that the forward and inverse problem are exactly related through a Möbius inversion. 

The chosen powerset mereology on the energy function restricts this approach to binary variables. However, categorical variables admit different mereologies with Möbius functions that lead to new notions of interaction no longer related to spin models \cite{beentjes2020higher,jansma2023higher}.

\subsubsection{Statistical mechanics}
In the approach above, one still needs access to observations of all microscopic variables to estimate the energy of observed states and solve the inverse problems. However, one could also start the line of reasoning from more macroscopic quantities, like averages over an ensemble. For example, one might only be able to measure the expected value of variables and their products, called correlation functions. For example, the two-point correlation function is given by
\begin{align}
    \langle X_1 X_2 \rangle = \sum_{x_1, x_2} p(X_1=x_1, X_2=x_2)x_1 x_2 
\end{align}
where the summation is over the full state space of the joint system $(X_1, X_2)$. The observed correlations will be the result of microscopic interactions, but the exact form of the interactions might be unknown. If we imagine a set $X$ of interacting particles moving through space, then particles might scatter together in groups, so the partition mereology $\Pi(X)$ is appropriate:
\begin{align}
    \langle X \rangle = \sum_{\pi \in \Pi(X)} u(\pi)
\end{align}
where $u(\pi)$ is the contribution by the partition $\pi$. For example, the 4-point correlation function can be decomposed into the following contributions, where variables that appear together in a given partition are connected by a line:
\input{corrFn.tex}
For example, a diagram like 
\begin{tikzpicture}[baseline=(current bounding box.mid),scale=0.4]
    \foreach \x/\y in {0/0, 1/0, 0/1, 1/1}
      \filldraw (\x, \y) circle (2pt);
    \draw (0,0) -- (0,1);
\end{tikzpicture}
corresponds to a term $u(\{X_1, X_3\}\{X_2\}\{X_4\})$, and might be interpreted as the contribution of the situation in which $X_1$ and $X_3$ interact, but $X_1$ and $X_4$ do not. The diagrammatic representation also makes clear that the bottom element $\{\{X_1\}\{X_3\}\{X_2\}\{X_4\}\}$ of the partition mereology corresponds to the non-interacting theory without any scattering. The correlation functions can be inverted over the partition mereology to give the contributions of individual diagrams, as well as those of terms involving a single $u$. For example, the first three orders of $u$ are easily seen to be given by
\begin{align}
    u(\{X_1\}) &= \langle X_1 \rangle \\
    u(\{X_1, X_2\}) &= \langle X_1 X_2 \rangle - \langle X_1 \rangle \langle X_2 \rangle \\
    u(\{X_1, X_2, X_3\}) &= \langle X_1 X_2 X_3 \rangle - \langle X_1 X_2 \rangle \langle X_3 \rangle - \langle X_1 X_3 \rangle \langle X_2 \rangle - \langle X_2 X_3 \rangle \langle X_1 \rangle + 2 \langle X_1 \rangle \langle X_2 \rangle \langle X_3 \rangle
\end{align}
These are precisely the well-known Ursell functions of statistical mechanics. While the first three Ursell functions coincide with the mixed central moments, the higher-order Ursell functions are different, and used throughout statistical mechanics. Ursell functions are related to the higher-order interactions of the mechanics and are essentially the cumulants of the theory, which is why they rely on exactly the same mereology of moments (see Appendix \ref{sec:stats}). Note, however, that Ursell functions are the partition-inverse of the moments, while the higher-order interactions of equilibrium dynamics are the powerset-inverse of the energy. 

\subsubsection{Quantum \& Statistical Field Theory}

In quantum and statistical field theory, random variables are replaced by field operators $\phi(x)$. The correlation functions---or \textit{Green's functions}---are then defined as expectation values of time-ordered products of field operators in the vacuum state $\ket{\Omega}$:
\begin{align}
    G^{(4)} (x_1, x_2, x_3, x_4) &= \bra{\Omega} T \phi(x_1) \phi(x_2) \phi(x_3) \phi(x_4) \ket{\Omega}\\
    &= \mathcal{Z}^{-1} \int \mathcal{D}\phi  e^{iS[\phi]} \phi(x_1) \phi(x_2) \phi(x_3) \phi(x_4)
\end{align}
where $T$ is the time-ordering operator, $S[\phi]$ is the action of the field theory, $\mathcal{Z} = \int \mathcal{D}\phi e^{iS[\phi]}$, and the integral is over all possible field configurations. Such time-ordered correlation functions are of central interest because they determine the particle scattering probabilities \cite{lehmann1955formulierung,peskin2018introduction} and thus form the macroscopic observables. 


As in the case of statistical classical mechanics, one can impose the partition mereology:
\begin{align}
    G^{(4)} (X) &= \sum_{\pi \in \Pi(X)} u(\pi)
\end{align}
This can of course be represented by the same diagrams as in the case of statistical mechanics in Equation \ref{eq:corrFnDiags}, but there is an important difference. By expanding $e^{iS[\phi]}$ as a power series around the non-interacting theory up to certain order and applying Wick's theorem to the resulting products of operators, one can show that the connected correlation functions are given by the sum of all Feynman diagrams with the same external vertices, but with an arbitrary number of internal vertices and loops. To emphasise that the diagrammatic representations of the decomposition may contain arbitrary internal processes, we draw the quantum diagrams with a shaded internal circle:
\vspace{1em}
\begin{center}
\begin{tikzpicture}[baseline=(current bounding box.mid),scale=1]
    \foreach \x/\y in {0/0, 1/0, 0/1, 1/1}
      \filldraw (\x, \y) circle (1pt);
    \draw (0,1) -- (1,0);
    \draw (0.5,0.5) -- (1, 1);
    \node at (0,1.5) {Statistical Mechanics:};
\end{tikzpicture} \hspace{4em}
\begin{tikzpicture}[baseline=(current bounding box.mid),scale=1]
    \foreach \x/\y in {0/0, 1/0, 0/1, 1/1}
      \filldraw (\x, \y) circle (1pt);

    \draw (0,1) -- (1,0);
    \draw (1, 1) -- (0.5, 0.5);
    \draw (0, 0) -- (0.5, 0);
    
    \filldraw[fill=gray] (0.5, 0.5) circle (5pt);
    \filldraw[fill=gray] (0.5, 0) circle (5pt);
    \node at (0,1.5) {Quantum Field Theory:};
\end{tikzpicture}
\vspace{1em}
\end{center}

A single connected component of one of the diagrams now represents an infinite sum over Feynman diagrams. For example, in a theory with quartic interactions:
\begin{align}
    \begin{tikzpicture}[baseline=(current bounding box.mid),scale=1]
        \filldraw (0, 0) circle (1pt);
        \filldraw (2, 0) circle (1pt);
        \draw (0,0) -- (2,0);
        \filldraw[fill=gray] (1, 0) circle (5pt);
    \end{tikzpicture}
    &=
    \begin{tikzpicture}[baseline=(current bounding box.mid),scale=1]
        \filldraw (0, 0) circle (1pt);
        \filldraw (2, 0) circle (1pt);
        \draw (0,0) -- (2,0);
    \end{tikzpicture}
    +
    \begin{tikzpicture}[baseline=(current bounding box.mid),scale=1]
        \filldraw (0, 0) circle (1pt);
        \filldraw (2, 0) circle (1pt);
        \filldraw (1, 0) circle (1pt);
        \draw (0,0) -- (2,0);
        \draw (1,0.2) circle (0.2);
    \end{tikzpicture}
    +
    \begin{tikzpicture}[baseline=(current bounding box.mid),scale=1]
        \filldraw (0, 0) circle (1pt);
        \filldraw (2, 0) circle (1pt);
        \filldraw (0.75, 0) circle (1pt);
        \filldraw (1.25, 0) circle (1pt);
        \draw (0,0) -- (2,0);
        \draw (1,0) circle (0.25);
    \end{tikzpicture}
    +
    \ldots
    +
    \begin{tikzpicture}[baseline=(current bounding box.mid),scale=1]
        \filldraw (0, 0) circle (1pt);
        \filldraw (2, 0) circle (1pt);
        \filldraw (1.5, 0) circle (1pt);
        \filldraw (0.5, 0) circle (1pt);
        \filldraw (1.5, 0.4) circle (1pt);
        \draw (0,0) -- (2,0);
        \draw (0.5,0.2) circle (0.2);
        \draw (1.5,0.2) circle (0.2);
        \draw (1.5,0.6) circle (0.2);
    \end{tikzpicture}
    + \ldots
\end{align}
A Möbius inversion over the partition mereology then gives an expression for the connected components of these graphs in terms of the correlation functions. For example:
\begin{align}
    \begin{tikzpicture}[baseline=(current bounding box.mid),scale=1]
        \filldraw (0, 0) circle (1pt);
        \filldraw (1, 0) circle (1pt);
        \draw (0,0) -- (1,0);
        \filldraw[fill=gray] (0.5, 0) circle (5pt);
    \end{tikzpicture}
    &= \bra{\Omega} T \phi(x_1) \phi(x_2)\ket{\Omega} - \bra{\Omega}\phi(x_1)\ket{\Omega}\bra{\Omega}\phi(x_2)\ket{\Omega} - 1\\
    \begin{tikzpicture}[baseline=(current bounding box.mid),scale=1]
        \filldraw (0, 1) circle (1pt);
        \filldraw (1, 1) circle (1pt);
        \filldraw (1, 0) circle (1pt);
        \draw (0,1) -- (0.5, 0.5);
        \draw (1,1) -- (0.5, 0.5);
        \draw (1,0) -- (0.5, 0.5);
        \filldraw[fill=gray] (0.5, 0.5) circle (5pt);
    \end{tikzpicture}
    &= \bra{\Omega} T \phi(x_1) \phi(x_2)\phi(x_3)\ket{\Omega} - \bra{\Omega}\phi(x_1)\ket{\Omega}\bra{\Omega}\phi(x_2)\phi(x_3)\ket{\Omega}\\
    &- \bra{\Omega}\phi(x_2)\ket{\Omega}\bra{\Omega}\phi(x_1)\phi(x_3)\ket{\Omega} - \bra{\Omega}\phi(x_3)\ket{\Omega}\bra{\Omega}\phi(x_1)\phi(x_2)\ket{\Omega}\\
    &+ 2\bra{\Omega}\phi(x_1)\ket{\Omega}\bra{\Omega}\phi(x_2)\ket{\Omega}\bra{\Omega}\phi(x_3)\ket{\Omega}
\end{align}

While this looks similar to the decomposition of the moments in Section \ref{sec:stats}, it should be emphasised that the correlation functions treated here are not the moments of some probability distribution over field configurations. Rather, the analogy holds only at the level of the expectation values, and the fact that one can define a moment generating function $\mathcal{Z}[J]$, as well as a cumulant generating function $\log(\mathcal{Z}[J])$. This fact has been used for decades to define cumulants in any setting where a suitable notion of \textit{average} can be defined, and is known as the generalised cumulant expansion method \cite{kubo1962generalized}. Such approaches have been used to define connected correlations in networks of neurons \cite{ocker2017linking}, on belief propagation graphs \cite{chertkov2006loop, welling2012cluster}, and in chemical dynamics \cite{domb1972cluster,klein1999chemical}.

\subsection{Chemistry \label{sec:chemistry}}
A significant challenge in chemistry is predicting the properties of molecules from their atomic configuration, commonly referred to as the quantitative structure-activity relationship (QSAR). To describe a molecular property $Q$ in terms of the configuration of a molecule $M$, one might associate a graph $G_M$ with the molecule, where the nodes are atoms and the edges are bonds. The property $Q$ can then be expressed as a sum over the subgraphs of $G_M$:
\begin{align}
    Q(M) = \sum_{G \leq G_M} q(G)
\end{align}
where $q(G)$ is the contribution of the subgraph $G$ to the property $Q$. A straightforward Möbius inversion over this subgraph mereology is of course possible, but some chemical properties are more naturally described by other mereologies. For example, a molecule's resonance energy is most naturally decomposed into contributions from acyclic graphs only. Motivated by such chemical questions, the authors of \cite{babic1996mobius} derived a closed-form expression for the Möbius function on this poset of acyclic graphs. 

Also studied in chemistry is how properties of a family of molecules are related. The authors of \cite{ivanciuc2005posetic}, for instance, study how toxic molecules from the family of chlorobenzenes are to guppies. A chlorobenzene is a benzene molecule where one or more hydrogen atoms have been replaced by chlorine atoms. That means that there is a certain partial order one can impose on the set of 13 possible chlorobenzenes (taking into account the 6-fold rotational symmetry of the benzene ring). Namely, two chlorobenzenes $c_1$ and $c_2$ are related by $c_1 \leq c_2$ if $c_2$ can be created from $c_1$ by adding chlorine atoms. The toxicity $T$ can then be expressed as a sum over the toxic contributions $t$ of the chlorobenzenes that come before it in the reaction chain:
\begin{align}
    T(c) = \sum_{c' \leq c} t(c')
\end{align}
Similarly, the authors of \cite{ivanciuc2007posetic} construct the poset of adding methyl groups to cyclobutanes. As the molecules higher-up in the partial order are constructed from the lower ones, one could expand the property of a molecule in terms of the contributions of the molecules that come before it. This means that every molecule defines its own poset and thus Möbius function, but these have been calculated and used to verify that higher-order contributions decrease so that truncating the expansion at a certain level leads to a good approximation of the property \cite{ivanciuc2007posetic}. In chemistry, this method is known as the posetic cluster expansion \cite{klein1986chemical}. Recently, the authors of \cite{barker2024multilevel} considered how a molecule's electronic potential decomposes either over atomic nuclei, or molecular fragments, and described similar truncations of Möbius inversions that can improve both the speed and accuracy of electronic potential calculations.

\subsection{Game Theory \label{sec:gameTheory}}
In coalitional game theory, players can form coalitions and cooperate, potentially increasing their expected payoff. The core idea is that value might add synergistically, or superadditively (for example: a pair of shoes might be worth more than twice the value of an individual shoe). Therefore, one could decompose the total value $v(S)$ of a coalition $S$ into the sum of the synergistic contributions $w(R)$ of all subsets of $S$:
\begin{align}
    v(S) = \sum_{R \subseteq S} w(R)
\end{align}
If the value of the coalitions is known, then this can be inverted over the powerset mereology to yield a definition of the synergistic contributions $w(R)$:
\begin{align}
    w(S) = \sum_{R \subseteq S} (-1)^{|S|-|R|} v(R)
\end{align}
Calculating the coalition synergy is of great practical interest because it allows the coalition payoff to be fairly distributed among its members. Since the synergy of a given coalition cannot be attributed to any single member, it should be distributed evenly among all members. The fair expected payout for any individual player $i$ in a coalition involving $N$ players is then the average synergy that player $i$ adds to each possible subcoalition that includes them:
\begin{align}
    \phi_i = \sum_{S \subseteq N: i \in S} \frac{w(S)}{|S|}
\end{align}
This is a very well-known quantity, known as the Shapley value for player $i$ \cite{shapley1953value}. It is more commonly written as
\begin{align}
    \phi_i = \frac{1}{|N|} \sum_{S \subseteq N\setminus\{i\}} {|N|-1 \choose{|S|}} (v(S\cup\{i\}) - v(S))
\end{align}
The Shapley value is thus the Möbius inverse of the normalised synergy over the subcoalition mereology. It is the unique payout function that satisfies such favourable and practical properties that its inventor was awarded the Nobel Prize in Economics. Choosing a different normalisation in the decomposition of $\phi_i$, one that for example depends on the identity of player $i$, an identical Möbius inversion recovers a family of distribution rules \cite{billot2005share}.

\subsection{Artificial Intelligence \label{sec:ml}}
Predictive machine learning models are generally trained on many features, and once the final model has been constructed it can be difficult to determine how much each feature contributes to the prediction. One way to address this issue is to decompose the prediction of the model into contributions of individual features, and using the Shapley value from the previous section to assign a total contribution to each feature, replacing the value function $v(S)$ with the model's prediction, and marginalising over the features not in $S$ \cite{cohen2007feature,williamson2020efficient,fryer2021shapley}. This allows one to determine which features are most important for a particular model's prediction, as well as which groups of features show synergistic effects.

In generative machine learning, the features do not contribute to a prediction, but to a probability distribution. Energy-based models like Restricted Boltzmann Machines, in which the probability of a sample is essentially the Boltzmann weight of a statistical physics model, have led to particularly fruitful insights into the relationship between a model's internal structure and the encoded distribution. An argument similar as in section \ref{sec:equilDyn} revealed that the Möbius inverse of the model's energy function over the powerset mereology captures the feature interactions, with higher-order interactions corresponding to synergistic effects among the features \cite{nguyen2017inverse,cossu2019machine,jansma2023-thesis,jansma2023higher}.

Even before the modern success of machine learning techniques, the study of artificial intelligence has inspired formal methods to model reasoning agents. Dempster-Shafer theory, or \textit{evidence} theory, is a generalisation of Bayesian probability theory \cite{dempster1967upper,shafer1976mathematical} (though its precise relationship to Bayesian inference is controversial \cite{dezert2013dempster}) that formalises how agents combine evidence to update beliefs. Given a system $S$, the set $U$ of possible states of $S$ is called a \textit{universe}. A proposition $p$ about the state of $S$ can be identified with the subset of $U$ in which $p$ is true. The set of all propositions therefore naturally admits a powerset mereology on $U$. Every proposition $A$ is then assigned a \textit{belief mass} $m(A)$ through the belief assignment function $m: \mathcal{P}(X) \to [0, 1]$, where $\sum_{T\in \mathcal{P}(U)} m(T) = 1$. The belief mass of $A$ is the belief added by observing $A$ that was not available upon observing a strict subset of $A$. The total belief an agent has about a proposition $A$ is then given by a sum over the powerset mereology on the universe:
\begin{align}
    Bel(A) = \sum_{B \subseteq A} m(B)
    \intertext{such that the belief assignment function can be expressed in terms of the belief of agents after observing varying evidence}
    m(A) = \sum_{B \subseteq A} (-1)^{|A|-|B|} Bel(B)
\end{align}
which is used to define the generalised Bayesian update rule. While historically Dempster-Shafer theory has used the powerset mereology, modern approaches have generalised this to other mereologies where properties of the Möbius function are known \cite{grabisch2009belief,zhou2013belief}. 

\section{Application: Decomposing the KL-divergence \label{sec:kl}}
To illustrate how the presented framework might be used to derive new measures of interactions, we introduce a novel decomposition of the Kullback-Leibler (KL) divergence. The KL-divergence is a measure of the difference between two probability distributions $p$ and $q$ and is of central importance in machine learning, statistics, and information geometry. It is defined as:
\begin{align}
    D_{KL}(p|q) &= \sum_{x} p(X=x) \log \frac{p(X=x)}{q(X=x)}\\
    &= \mathbb{E}_{x \sim p} \left[ \log \frac{p(x)}{q(x)} \right]
\end{align}
In general, $p$ and $q$ are multivariate distributions and $X = \{X_1, X_2, \ldots X_n\}$, so the sum is over the entire state multivariate state space $\mathcal{X}$ of $X$. The KL-divergence assigns a number to the average difference between $p$ and $q$, but does not tell you where this difference comes from---in this sense it is a `macroscopic' or `global' property of the two distributions. A large KL-divergence could be the result of a small difference in the mean of each variable, or a large difference among only a few of the higher-order moments. Disentangling this information could yield useful and actionable insights, for example when the KL-divergence is used as a training loss in an optimisation problem. Knowing where the discrepancy between the target and model distribution comes from can help in understanding the model's behaviour and improving it. 

Let us assume that the ordering of the variables does not matter and impose the powerset mereology. Let $D_{KL}(p|q; X)$ be the KL-divergence between two multivariate distributions $p$ and $q$ over a set $X$ of $n$ variables. We can decompose the KL-divergence into the sum of `microscopic' contributions $\Delta_{p|q}$ of all subsets of $X$:
\begin{align}
    D_{KL}(p|q; X) &= \sum_{S \subseteq X} \Delta_{p|q}(S)\\
    \nonumber &\Updownarrow\\
    \Delta_{p|q}(X) &= \sum_{S \subseteq X} (-1)^{|X|-|S|} D_{KL}(p|q; S)
\end{align}
To make sense of this formula, we have to decide on a definition of $D_{KL}(p|q; T)$ when $T \subseteq X$. One obvious choice is to define it as the KL-divergence between the marginal distributions $p_T$ and $q_T$ where all variables outside of $T$ are marginalised over. 
\begin{align}
    D_{KL}(p|q; T) = D_{KL}(p_T|q_T)
\end{align}
where $p_T(T=t) = \sum_{x} p(X\setminus T=x, T=t)$ and similarly for $q_T$. For example, when $X = \{X_1, X_2, X_3\}$, the third-order contribution to the KL-divergence is given by
\begin{align}
    \nonumber \Delta_{p|q}(X_1, X_2, X_3) &= D_{KL}(p|q) - D_{KL}(p_{X_1, X_2}|q_{X_1, X_2}) - D_{KL}(p_{X_1, X_3}|q_{X_1, X_3})\\
    &- D_{KL}(p_{X_2, X_3}|q_{X_2, X_3}) + D_{KL}(p_{X_1}|q_{X_1}) + D_{KL}(p_{X_2}|q_{X_2}) + D_{KL}(p_{X_3}|q_{X_3})
\end{align}

\begin{figure}
    \includegraphics[width=0.3\textwidth]{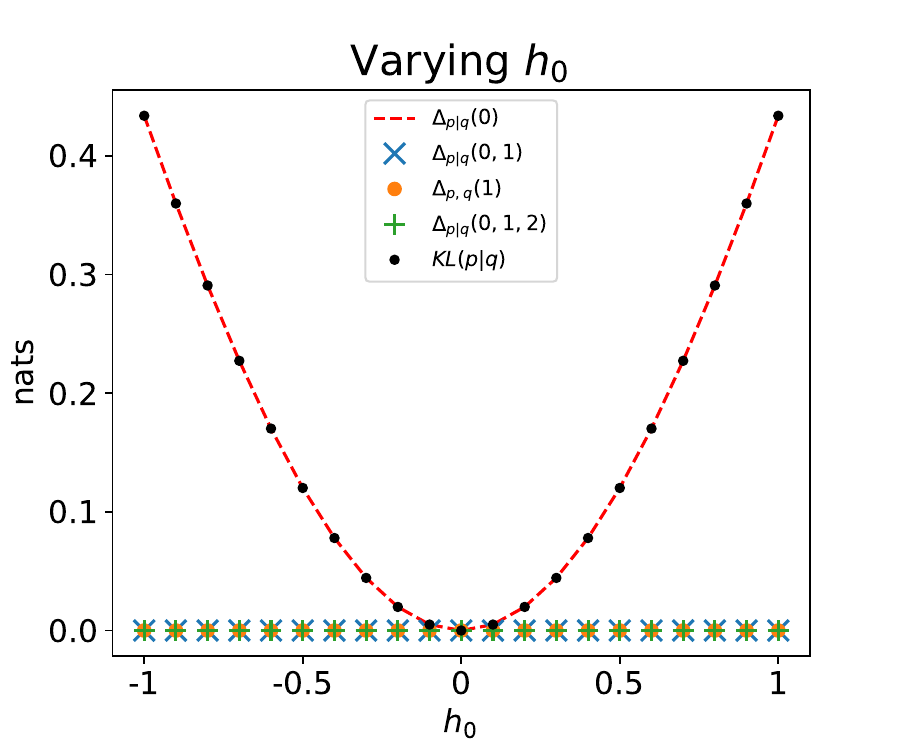}
    \includegraphics[width=0.3\textwidth]{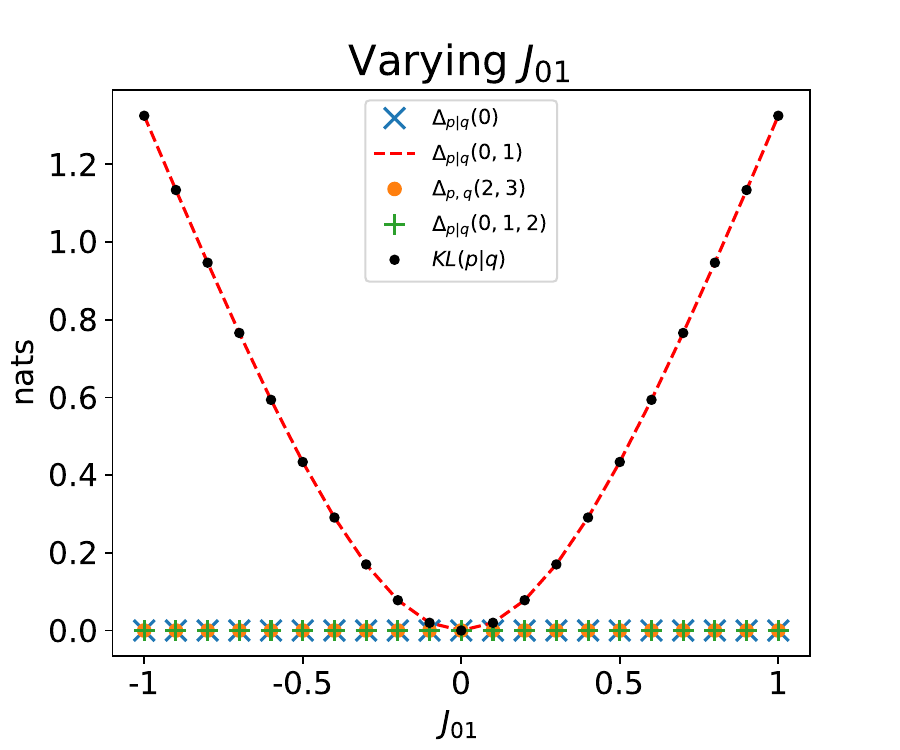}
    \includegraphics[width=0.289\textwidth]{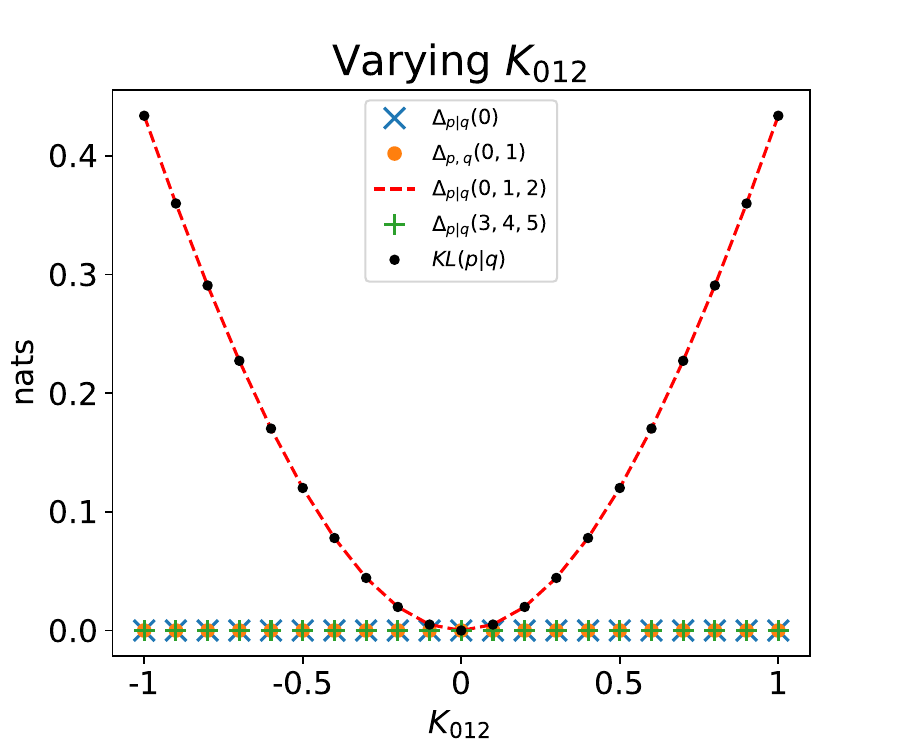}
    \caption{The KL-divergence between the uniform distribution and an interaction Ising model ($n=4$) grows with the interaction strength, but the decomposition reveals that this growth is not uniform across all subsets of variables. In fact, the decomposition is able to exactly identify the set of variables that is responsible for the deviation from the uniform distribution.\label{fig:KLdecomp}}
\end{figure}

To investigate if the $\Delta_{p|q}$ measure indeed separates different contributions to the KL-divergence, we calculate the KL-divergence between a reference distribution and various Ising models with up to 3-point interactions. Consider the following Ising distribution:
\begin{align}
    p(x; h, J, K) = \frac{1}{Z} \exp\left(\sum_{i} h_i x_i + \sum_{i\neq j} J_{ij} x_i x_j + \sum_{i\neq j\neq k} K_{ijk} x_i x_j x_k\right)
\end{align}
where the spins $x_i \in \{-1, 1\}$. The uniform distribution corresponds to an Ising model $p(x; 0, 0, 0)$ with $h_i = J_{ij} = K_{ijk} = 0$. We compare $p(x; 0, 0, 0)$ to Ising models with only an external field, only pairwise nearest-neighbour coupling, or only 3-point interactions. To do so, define $h^{(0)}$, $J^{(0)}$, and $K^{(0)}$ as:
\begin{align}
    h^{(0)}_i = h_0 \delta^i_0 \quad J^{(0)}_{ij} = J_{01} \delta^i_0\delta^j_1 \quad K^{(0)}_{ijk} = K_{012} \delta^i_0\delta^j_1\delta^k_2
\end{align}
where $\delta^i_j$ is the Kronecker delta. The corresponding Ising distributions are given by
\begin{align}
    q_1(x) & \coloneq p(x; h^{(0)}, 0, 0) = Z^{-1} \exp\left(h_0 x_0\right)\\
    q_2(x) & \coloneq p(x; 0, J^{(0)}, 0) = Z^{-1} \exp\left(J_{01} x_0 x_1\right)\\
    q_3(x) & \coloneq p(x; 0, 0, K^{(0)}) = Z^{-1} \exp\left(K_{012} x_0 x_1 x_2\right)
\end{align}
In each of these three interacting scenarios, we vary the strength of the non-zero coupling from $-1$ to $1$, and calculate the KL-divergence $D_{KL}\left(p(x;0, 0, 0)|q_i(x)\right)$, as well as its decomposition into $\Delta_{p|q}$ terms. The results for Ising models on 4 variables are shown in Figure \ref{fig:KLdecomp} (an associated Jupyter notebook is available at \url{https://github.com/AJnsm/KLdecomposition}). As expected, the KL-divergence grows with the absolute coupling strength, but the $\Delta_{p|q}$'s offer more information and precisely disentangle the various contributions. For example, when varying $K_{012}$, all $\Delta_{p|q}$ at orders 1 and 2 are zero, and only $\Delta_{p|q}(012)$ is non-zero among the 3rd-order terms. Similar disentangling can be seen when varying 1- or 2-point couplings. This shows that the decomposition is able to precisely identify the variables that are responsible for the deviation from the uniform distribution. 

\begin{figure}
    \includegraphics[width=0.3\textwidth]{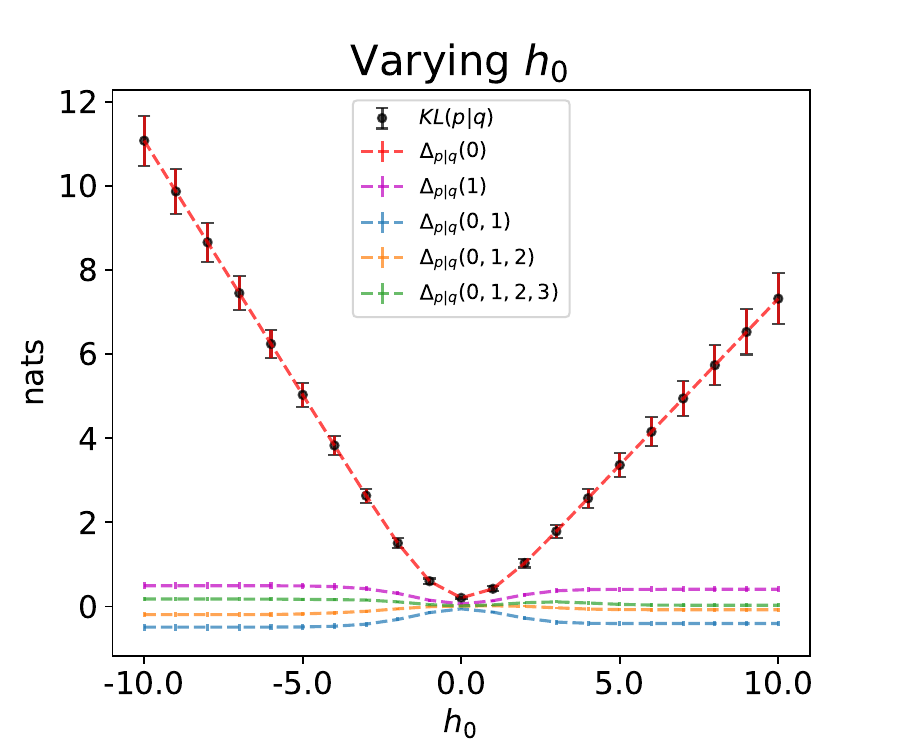}
    \includegraphics[width=0.3\textwidth]{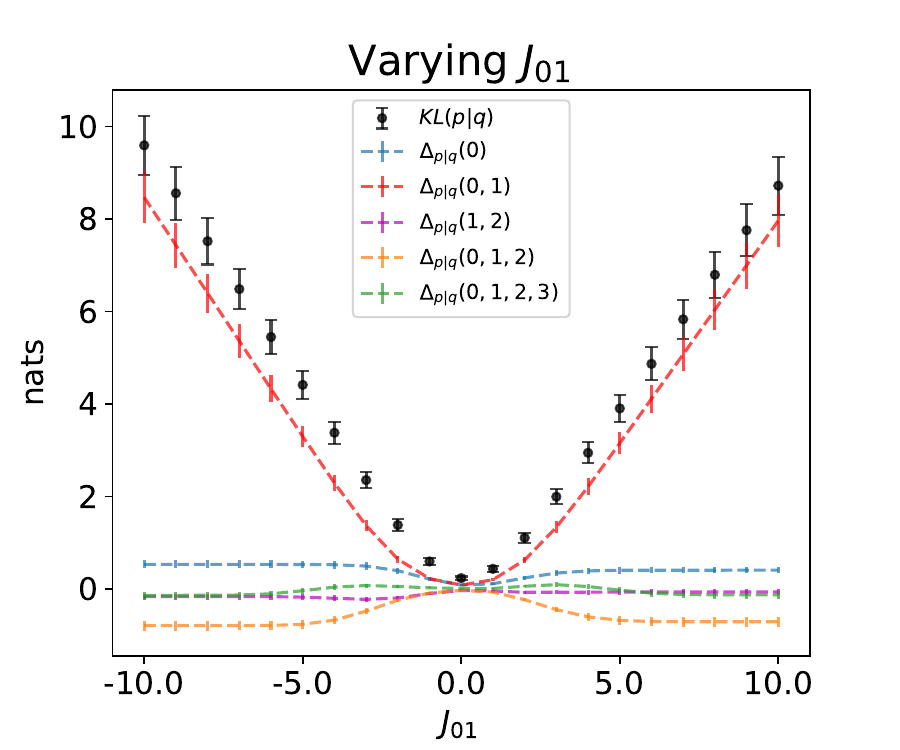}
    \includegraphics[width=0.289\textwidth]{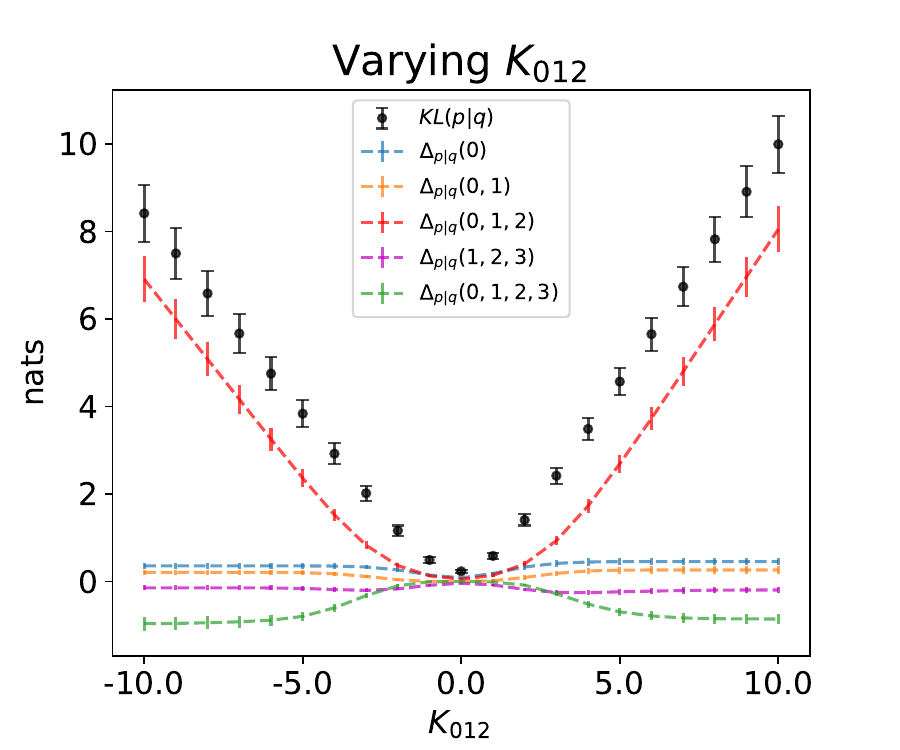}
    \caption{Decomposing the KL-divergence between two interacting Ising models ($n=4$) also indicates which variables are responsible for the discrepancy, though at weak coupling this identification becomes harder, especially for higher-order discrepancies. Errors bars are standard errors on the mean across 100 initialisations of randomised interacting reference distributions. \label{fig:KLdecomp_int}}
\end{figure}

In practice, the reference distribution will generally not be uniform, since one often compares an empirical distribution to a model distribution---neither of which should \textit{a priori} be uniform. To validate the behaviour of the $\Delta_{p|q}$'s in a more realistic scenario, we also calculated and decomposed the KL-divergence between two interacting Ising models. In this case, the reference distributions are those with random couplings $h^{(r)}$, $J^{(r)}$, and $K^{(r)}$:
\begin{align}
    h^{(r)}_i \sim \mathcal{N}(0, 1) \quad J^{(r)}_{ij} \sim \mathcal{N}(0, 1)\delta^{(j-i)\text{mod} N}_1 \quad K^{(r)}_{ijk} \sim \mathcal{N}(0, 1) (1 - \delta^j_i)(1 - \delta^k_i)(1 - \delta^k_j)
\end{align}
where for each set of indices, a single sample is drawn from the standard normal, which is then assigned to all permutations of the indices to enforce symmetry of $J$ and $K$. The reference distribution is then given by $p(x;h^{(r)}, J^{(r)}, K^{(r)})$. The other distribution $q_i(x)$ is a copy of this distribution, but with one of the $i$th-order couplings varying from $-10$ to $10$ (namely, $h_0$, $J_{01}$, or $K_{012}$, see Figure \ref{fig:KLdecomp_int}). We then calculate the KL-divergence between the interacting reference distribution $p(x;h^{(r)}, J^{(r)}, K^{(r)})$ and each $q_i$. The results across 100 initialisations of the randomised reference distribution are shown in Figure \ref{fig:KLdecomp_int}. The largest $\Delta_{p|q}$ still clearly identify the set of variables that is responsible for the discrepancy, but at weak coupling---that is, small discrepancy---small but significantly non-zero $\Delta_{p|q}$ among other variables make identifying the true source of the discrepancy more difficult.

Disentangling the source of a nonzero KL-divergence could allow computational resources to be spent more efficiently when optimising a model with respect to an empiricial distribution. In addition, the KL-divergence has been used to quantify the effects of evolutionary pressure, by comparing the distribution of selected vs neutral pheno- and genotypes \cite{hledik2022accumulation}. Disentangling the KL-divergence among these features would show precisely where the evolutionary pressure is acting, and could lead to a better understanding of the evolutionary process.

\section{Application: Renormalisation \label{sec:renormalisation}}
The presented framework can also be used to derive the effect of coarse-grainings, or renormalisation group transformations. The renormalisation group is a method to study the behaviour of a system at different scales. By grouping variables together, new `coarse-grained' variables emerge, and one can study how the new variables interact. Our framework provides a very general language in which to describe the effect on microscopic interactions of coarse-graining a system, since the microscopic interactions are defined relative to a chosen mereology, and a coarse-graining is essentially a modification of a system's mereology.

In this section, we will describe coarse-grainings in terms of partially ordered sets, and how this constrains the new Möbius function. We will then use this to derive the renormalised couplings of the 1D Ising model, without ever having to write down or sum over a partition function. The content of this section can be summarised as ``renormalised interactions are Möbius inversions over a Galois connection".

\subsection{Coarse-grainings are Galois connections}
Given a system $S$ with mereology $(\mathcal{M}_S, \leq)$, a coarse-graining is a function $\rho: (\mathcal{M}_S, \leq) \to (\mathcal{M}_{\tilde{S}}, \preceq)$ that maps parts of $S$ to parts of a coarse-grained system $\tilde{S}$ and for which $|\mathcal{M}_{\tilde{S}}| \leq |\mathcal{M}_{S}|$. One can decompose macroscopic variables before and after the coarse graining separately:
\begin{align}
    Q(b) = \sum_{a \leq b} q(a) \quad \text{ and }\quad  \tilde{Q}(\tilde{b}) = \sum_{\tilde{a} \preceq \tilde{b}} \tilde{q}(\tilde{a})
\end{align}
In words: coarse graining a system is a change in mereology, together with a prescription to modify $Q$. This in turn defines the renormalised couplings $\tilde{q}$:
\begin{align}
    \tilde{q}(\tilde{a}) = \sum_{\tilde{a} \preceq \tilde{b}} \mu_{\mathcal{M}_{\tilde{S}}}(a, b) \tilde{Q}(b)
\end{align}
When studying the renormalisation group, one is usually interested in relating renormalised couplings to the original couplings, so we want to find a way to relate the two mereologies, their Möbius functions, and the macroscopic quantities. We claim that in most cases of practical interest, the two mereologies $\mathcal{M}_S$ and $\mathcal{M}_{\tilde{S}}$ are both lattices. The most obvious example is that of two powerset mereologies $(\mathcal{P}(S), \subseteq)$ and $(\mathcal{P}(\tilde{S}), \subseteq)$, where the coarse-graining simply forgets a set of variables $U\subset S$ to obtain $\tilde{S} = S\setminus U$. This is usually called \textit{decimation}, and can be captured by a coarse-graining function:
\begin{align}
    \rho: \mathcal{P}(S) &\to \mathcal{P}(\tilde{S}) \label{eq:coarse_graining_powerset}\\
    A &\mapsto A\setminus U
\end{align}
Recall that a (covariant) Galois connection between two partial orders $P$ and $Q$ is given by a pair of monotone functions $f: P \to Q$ and $g: Q \to P$ such that for all $p \in P$ and $q \in Q$:
\begin{align}
    p \leq g(f(p)) \quad \text{and} \quad f(g(q)) \leq q
\end{align}
If this is the case, then we call $f$ the left adjoint, and $g$ the right adjoint, written as $f \dashv g$.
Now define the following two functions. 
\begin{align}
    \sigma: \mathcal{P}(\tilde{S}) &\to \mathcal{P}(S)\\
    A &\mapsto A \cup U\\
    \tau: \mathcal{P}(\tilde{S}) &\xhookrightarrow{} \mathcal{P}(S)\\
    A &\mapsto A 
\end{align}
Then it can be verified that $\rho \dashv \sigma$ and $\tau \dashv \rho$, so it is both the left adjoint of the union map, and the right adjoint of the inclusion map (and accordingly preserves both upper and lower bounds). More generally, when the mereologies form a lattice, given any coarse graining map $\rho$, we can construct a right adjoint $\sigma$ by $\sigma(b) = \bigvee \{a|\rho(a) \leq b\}$. 

This is a useful observation in light of Rota's Galois connection theorem:
\begin{theorem}[Rota's Galois Connection Theorem]
Let $\rho: P \to Q$ and $\sigma: Q \to P$ be a Galois connection $\rho\dashv \sigma$. Then
\begin{align}
    \sum_{\substack{u \in P \\ \rho(u) = y}} \mu_P(x, u) = \sum_{\substack{v \in Q \\ \sigma(v) = x}} \mu_Q(v, y) \label{eq:RGCT}
\end{align}
\end{theorem}
If the right hand side of Equation \eqref{eq:RGCT} contains only a single term, then this gives a way to evaluate the Möbius function on the renormalised mereology in terms of the original one. Since $\sigma$ is injective, this is necessarily the case, and we can write:
\begin{align}
    \mu_{\mathcal{M}_{\tilde{S}}}(\tilde{a}, \tilde{b}) = \sum_{\substack{u \in \mathcal{M}_{S} \\ \rho(u) = b}} \mu_{\mathcal{M}_S}(\sigma(a), b)
\end{align}
Leading to the following expression for the renormalised interactions:
\begin{align}
    \tilde{q}(\tilde{b}) = \sum_{\tilde{a} \leq \tilde{b}}\left(\sum_{\substack{u \in \mathcal{M}_S \\ \rho(u) = \tilde{b}}} \mu_{\mathcal{M}_S}(\sigma(\tilde{a}), u)\right) \tilde{Q}(\tilde{a})
\end{align}

\begin{figure}
\[\begin{tikzcd}
	&& {\{{a, b, c}\}} \\
	&&&&&&&&&&& {\{a, c\}} \\
	{\{a, b\}} && {\{b, c\}} && {\{a, c\}} \\
	&&&&&&&&& {\{a\}} &&&& {\{c\}} \\
	{\{a\}} && {\{b\}} && {\{c\}} \\
	&&&&&&&&&&& \emptyset \\
	&& \emptyset
	\arrow[color={rgb,255:red,92;green,214;blue,92}, from=1-3, to=2-12]
	\arrow[color={rgb,255:red,214;green,92;blue,92}, curve={height=12pt}, from=2-12, to=1-3]
	\arrow[from=2-12, to=4-10]
	\arrow[from=2-12, to=4-14]
	\arrow[from=3-1, to=1-3]
	\arrow[color={rgb,255:red,92;green,214;blue,92}, curve={height=18pt}, from=3-1, to=4-10]
	\arrow[from=3-3, to=1-3]
	\arrow[color={rgb,255:red,92;green,214;blue,92}, from=3-3, to=4-14]
	\arrow[from=3-5, to=1-3]
	\arrow[color={rgb,255:red,92;green,214;blue,92}, curve={height=-12pt}, from=3-5, to=2-12]
	\arrow[color={rgb,255:red,214;green,92;blue,92}, curve={height=30pt}, from=4-10, to=3-1]
	\arrow[from=4-10, to=6-12]
	\arrow[color={rgb,255:red,214;green,92;blue,92}, curve={height=24pt}, from=4-14, to=3-3]
	\arrow[from=4-14, to=6-12]
	\arrow[from=5-1, to=3-1]
	\arrow[from=5-1, to=3-3]
	\arrow[color={rgb,255:red,92;green,214;blue,92}, curve={height=6pt}, from=5-1, to=4-10]
	\arrow[from=5-3, to=3-1]
	\arrow[from=5-3, to=3-5]
	\arrow[color={rgb,255:red,92;green,214;blue,92}, curve={height=12pt}, from=5-3, to=6-12]
	\arrow[from=5-5, to=3-3]
	\arrow[from=5-5, to=3-5]
	\arrow[color={rgb,255:red,92;green,214;blue,92}, curve={height=18pt}, from=5-5, to=4-14]
	\arrow[color={rgb,255:red,214;green,92;blue,92}, from=6-12, to=5-3]
	\arrow[from=7-3, to=5-1]
	\arrow[from=7-3, to=5-3]
	\arrow[from=7-3, to=5-5]
	\arrow[color={rgb,255:red,92;green,214;blue,92}, curve={height=12pt}, from=7-3, to=6-12]
\end{tikzcd}\]
\caption{The lattice of subsets of $\{a, b, c\}$, with the Galois connection $\rho \dashv \sigma$ between the powerset mereologies of $\{a, b, c\}$ and $\{a, c\}$. The coarse graining $\rho$ (shown in green) is given by $\rho(A) = A\setminus \{b\}$, and the right adjoint $\sigma$ (shown in red) is given by $\sigma(B) = B \cup \{b\}$. The Möbius function on the right mereology can be calculated from the Möbius function on the left mereology using Rota's Galois connection theorem. One could view this as an instance of a free-forgetful, or inclusion-restriction adjunction. \label{fig:GaloisConnection}}
\end{figure}

\subsection{A New Derivation of Renormalised Ising Couplings}
Let us now apply this to a situation where the exact solution in well-known: the renormalised couplings of the 1D Ising model under decimation of the even spins. The decimation procedure corresponds to the Galois connection $\rho \dashv \sigma$, where $\rho$ is given by Equation \eqref{eq:coarse_graining_powerset} and sigma is its right adjoint (an example on three variables is shown in Figure \ref{fig:GaloisConnection}). Rota's Galois connection theorem (or the observation that $\mathcal{M}_S$ is still a Boolean algebra) then gives:
\begin{align}
    \mu_{\mathcal{M}_{\tilde{S}}}(\tilde{a}, \tilde{b}) &= \sum_{\substack{u \in \mathcal{M}_S \\ \rho(u) = \tilde{b}}} \mu_{\mathcal{M}_S}(\sigma(\tilde{a}), u) \\
    &= \begin{cases}
        (-1)^{|\tilde{a}| - |\tilde{b}|} & \text{if $\tilde{a} \subseteq \tilde{b}$ }\\
        0 & \text{otherwise}
    \end{cases}
\end{align}
This only leaves $\tilde{Q}(\tilde{a})$ to be determined. Before decimation, the macroscopic observables were given by $Q(b) = \log(p(b=1, S\setminus b = 0))$. Let us define the new observables by simply marginalising over the states of the decimated variables $R$, as this is the probabilistic equivalent of decimation:
\begin{align}
    \tilde{Q}(\tilde{b}) = \log\left(\sum_{r \in \{0, 1\}^{|R|}} p(b=1, R=r, S\setminus (b\cup R) = 0)\right)
\end{align}

Consider a 1D Ising model with $\{0, 1\}$ spins, and decimating every other spin. It can be readily verified that the renormalised pairwise coupling as a function of the original coupling $J$ and external field $h$ is given by:
\begin{align}
    \tilde{J}(J, h)  &= - \log(e^{-h} + e^{-2J-2h}) + 2 \log (e^{-h/2}+ e^{-J - \frac{3}{2}h}) - \log(1+e^{-h})\label{eq:renormJ_01}
\end{align}

Using the notation from Equation \eqref{eq:ising_3point}, we can write the pairwise interactions before and after decimation as:
\begin{align}
    I_{ab} &= - \log \frac{p_{110} p_{000}}{p_{100}p_{010}} \coloneq J\\
    \tilde{I}_{ac} &= -\log \frac{(p_{101} + p_{111}) (p_{000} + p_{010})}{(p_{100} + p_{110}) (p_{001} + p_{011})} \coloneq \tilde{J}
\end{align}
We then impose the specific symmetries of a 1D homogeneous Ising model with periodic boundary conditions and no external field:
\begin{align}
    p_{100} = p_{010} = p_{001} \coloneq a &\text{\quad (translation invariance)}\\
    p_{110} = p_{011} \coloneq b &\text{\quad(translation invariance)}\\
    p_{101} = p_{000} = p_{100} = a &\text{\quad(no field)}\\
    p_{111} \coloneq c
\end{align}
If we in addition demand that that only nearest neighbours interact, then $p_{111} = b^2$ by translation invariance. Let us set $a=1$ (since any normalisation constant would cancel in the fraction). That makes the renormalised coupling equal to
\begin{align}
    \tilde{J} = - \log \frac{2(1+b^2)}{(1+b)^2}
    \intertext{substituting $J = -\log b$, this leads to the recursion formula for the renormalised coupling:}
    \tilde{J} = \log \left(\frac{1}{2} \text{sech}(J) + \frac{1}{2}\right)
\end{align}

Note that this has a fixed point only at $J=0$, which is different from the $\pm1$ Ising model with fixed points at $J=0$ and $J=\infty$. This is allowed because the fieldless $\pm$ Ising model is equivalent only to a $\{0, 1\}$ Ising model with a field.  

Similarly, one can allow for the possibility where $p_{000}\neq p_{100}\coloneq a$, which captures the effect of adding a symmetry breaking field. Under this assumption, one can keep track of the extent to which the states break this symmetry, and define
\begin{align}
    p_{100} = p_{010} = p_{001} \coloneq a &\text{\quad (translation invariance)}\\
    p_{110} = p_{011} \coloneq a^2 b &\text{\quad(translation invariance, NN interaction)}\\
    p_{101} = a^2 &\text{\quad(symmetry breaking field)}\\
    p_{111} = a^3 b^2 &\text{\quad(only NN interaction)}
\end{align}
so that
\begin{align}
    \tilde{J} = - \log \frac{(a^2 + a^3b)(1+a)}{(a+a^2b)^2}
\end{align}
which, upon setting $a = \log (-h)$ and $b = \log(-J)$ indeed recovers the derived form of the renormalised Ising coupling in Equation \eqref{eq:renormJ_01}. Note that this approach yields the recursion relation for the renormalised couplings without ever requiring the Boltzmann distribution.

\section{Discussion \label{sec:discussion}}

\begin{table}[ht]
\centering
{\small{
\begin{tabular}{|r||l|l|l|}
\hline
Field of Study & Macro Quantity & Mereology & Micro Quantity/Interactions \\
\hline
\hline
Statistics  & Moments & Powerset & Central moments \\
            & Moments & Partitions & Cumulants \\
            & Free moments & Non-crossing partitions & Free cumulants \\
            & Path signature moments & Ordered partitions & Path signature cumulants \\
\hline
Information Theory & Entropy & Powerset & Mutual information \\
            & Surprisal & Powerset & Pointwise mutual information \\
            & Joint Surprisal & Powerset & Conditional interactions \\
            & Mutual Information & Antichains & Synergy/redundancy atoms\\
\hline
Biology & Pheno- \& Genotype & Powerset & Epistasis \\
        & Gene expression profile & Powerset & Genetic interactions \\
        & Population statistics & Powerset & Synergistic treatment effects\\
\hline
Physics & Ensemble energies & Powerset & Ising interactions \\
        & Correlation functions & Partitions & Ursell functions \\
        & Quantum corr. functions & Partitions & Scattering amplitudes \\
        \hline
Chemistry & Molecular property & Subgraphs & Fragment contributions \\
        & Molecular property & Reaction poset & Cluster contributions \\
\hline
Game Theory & Coalition value & Powerset & Coalition synergy\\
            & Shapley value & Powerset & Normalised coalition synergy\\
\hline
Artificial Intelligence & Generative model probabilities & Powerset & Feature interactions\\
                & Predictive model predictions & Powerset & Feature contributions\\
                & Dempster-Shafer Belief & Distributive & Evidence weight\\
                & $D_{KL}(p|q)$ & Powerset & $\Delta_{p|q}$ (See Sec. \ref{sec:kl})\\
\hline
\end{tabular}}}
\caption{An overview of the various ways in which macroscopic quantities can be linked to microscopic interactions by the Möbius inversion associated with a certain decomposition.}
\label{tab:summaryTable}
\end{table}

The aim of this study has been to provide a unified perspective on the various notions of higher-order interactions in complex systems. We have shown that decomposing a system into a mereological structure leads to a unique definition of the interactions among the parts, through a Möbius inversion of the outcome of interest. Furthermore, this presented a precise meaning of the word \textit{higher-order}, namely higher with respect to the partial order of the mereology. In particular, the smallest elements of each of the studied mereologies corresponded to the `reductionistic' parts of the theory. We found that this approach reproduces well-known notions of higher-order structure in a variety of scientific fields, an overview of which is provided in Table \ref{tab:summaryTable}. While some of the relationships in Table \ref{tab:summaryTable} have previously been described as Möbius inversions, to our knowledge this is the first time that the shared structure underlying these definitions has been made explicit. In addition, the framework lays out an approach for defining new kinds of microscopic interactions relative to a macroscopic quantity and its mereology. To illustrate this in practice, we used our framework to derive a new decomposition of the KL-divergence that allows for the identification of the variables that are responsible for the discrepancy between two distributions. We further showed that describing systems at the level of their mereology allows for efficient calculations of renormalised interactions, using the 1D Ising model as an example.

It should be noted that the Möbius inversion theorem can also be scientifically practical beyond its use in the mereological framework presented here. For example, it has been used by the author of \cite{chen1990modified} to derive an expression for phonon densities in a crystal lattice, but in the context of a number-theoretic trick rather than a decomposition of the system. Similarly, the inclusion-exclusion principle is ubiquitous, but not always indicative of an instance of the presented framework. In this study, our aim has been to explicitly show how the Möbius inversion theorem defines interactions relative to a chosen mereology and macroscopic observables, not to give a review of all the ways in which the theorem can be applied. 

As a general rule, observations of a system provide access to the macroscopic features (moments, phenotypes, energies, predictions, \textit{etc}), which can then be used to infer the microscopic interactions (cumulants, genetic interactions, Ising interactions, feature contributions, \textit{etc}), so the Möbius inversion theorem can offer a solution to the inverse problem and quantify emergence by revealing to which extent the atomic contributions at the bottom of the mereology are insufficient to explain macroscopic properties. However, such a statement must be accompanied by two caveats. First, note that a Möbius inversion only solves the inverse problem relative to a chosen mereology. While there are no universal criteria for a `good' decomposition, the mereology should be motivated by knowledge of the system's structure and the macroscopic property of interest. For example, a partition-based mereology may be natural when the macroscopic property depends on how different parts come together, while the powerset mereology is more suited to properties that depend on binary configurations of the parts. When there is a natural structure to the parts, like in the case of noncommuting variables or chemical structures, a decomposition that respects this structure, like non-crossing partitions or subgraphs, can be more appropriate. If the chosen mereology is not appropriate, then a Möbius inversion will not lead to meaningful microscopic interactions either. One example of this is the Möbius inversion on the partition mereology that related scattering amplitudes of quantum field theory to correlation functions. The Möbius inversion still gave infinite sums over Feynman diagrams, not the contribution of individual diagrams, so the interesting microscopic interactions (connected parts of individual diagrams) were not directly accessible from the macroscopic observables. Furthermore, collections of particles can carry more structure, like fermion number and colour, which also implies that partitions are not the most appropriate decomposition for the problem. The microscopic interactions therefore inherit their justification from the mereology, and are not guaranteed to be `real' or useful.

As a second caveat, note that some canonic examples of emergence cannot be studied with the presented framework, because the Möbius inversion is only possible if the macroscopic quantity is defined on every element of the mereology. For many classic examples of emergence, like bird flocks, temperature, wetness \textit{etc}, this is not possible (a single bird does not flock, and a single water molecule has no temperature or wetness). An interesting example of this problem arose in the partial information decomposition. Under the antichain mereology, estimating the contribution of the information atoms is only possible with a suitable definition of redundant information on each of the antichains. Much of the PID literature has focused on resolving this ambiguity in different ways, but no consensus has been reached thus far. However, this is also a reflection of the versatility of the Möbius inversion framework: even when the decomposition results in ambiguity, different resolutions of this ambiguity can give rise to a rich set of higher-order structure, which in the case of the PID have been used to characterise different properties of neural information processing \cite{wibral2017quantifying,sherrill2021partial,luppi2022synergistic}.

A particularly interesting phenomenon not explored in the present study is that of order dualities. As was already observed in \cite{jansma2023higher}, inverting a lattice leads to dual notions of higher-order interactions. In information theory, the dual to mutual information was found to be conditional mutual information, and a similar dual was derived for Ising interactions. It is an unexplored but interesting question whether the quantities dual to the ones presented here also offer a meaningful interpretation. It is clear that the rooted mereologies used in this study can be inverted to yield new mereologies, but deriving the corresponding dual interactions is left for future work.

Another exciting possibility is using the presented framework to transfer insights from one scientific discipline to another. For example, in the so-called \emph{cluster variation method}, physical intuition has motivated the truncation of sums over lattice subsets to obtain approximations to thermodynamic quantities of crystals \cite{kikuchi1951theory}. This summation can be seen as a truncated Möbius inversion over the lattice of physical crystal lattice sites \cite{an1988note, morita1990cluster}, and could be explored as an approximation scheme in other settings as well. In fact, precisely such a truncation has been suggested in decompositions of chemical properties of molecules \cite{ivanciuc2007posetic}. Whether it is useful to describe partial sums over scattering diagrams similarly as truncated Möbius inversions is---to our best knowledge---an open question. In addition, the authors of \cite{jansma2023high} use causal discovery methods to improve the estimation of genetic interactions (the Möbius inverse of gene-expression profiles). Since the estimation of microscopic interactions from macroscopic observables can require many observations, similar methods might be able to improve the estimability of higher-order interactions in other fields.

Further future work could explore different kinds of mereologies and the higher-order structure they imply. For example, a recent study has started to explore the combinatorics of nucleotide sequences in polyploid genomes by decomposing sequencing data over the lattice of integer partitions ordered by refinement \cite{ranallo2020genomescope}, and it has long been noted that the secondary structure of RNA molecules can be described in terms of non-crossing partitions \cite{penner1993spaces}. Some special mereologies turned out to be associated to famous constructions: posets with only a single nontrivial level correspond to `reductionistic' theories, and Galois connections turned out to describe coarse-grainings. One might similarly suspect that mean-field approximations can be identified with a certain class of mereologies. More speculatively, mereologies beyond locally finite partial orders could be explored by suitably generalising the Möbius inversion theorem. It is well-known that the Möbius inversion theorem can be generalised to a more general class of skeletal categories that includes not just posets but also monoids and groupoids \cite{leroux1975les,content1980categories,haigh1980mobius,leinster2012notions}, as well as to bialgebras \cite{kock2020mobius}. Whether these more general decompositions faithfully and fruitfully describe higher-order structure in complex systems is a mostly unexplored and open question. We hope that this work inspires others to explore novel mereologies and discover new types of higher-order structures in complex systems.

\section*{Acknowledgements}
The connections to game theory, chemistry, and path statistics were the result of fruitful discussions with Jürgen Jost, Guillermo Restrepo, and Darrick Lee, respectively. The author is also grateful to Eckehard Olbrich and Fernando Rosas for insightful discussions on the partial information decomposition, as well as to Hadleigh Frost for his comments on diagrammatic sums in quantum field theory, and Joachim Kock for his comments on mereology. 

\appendix

\section{Möbius Inversions and Mereology in Statistics \label{sec:stats}}

Of central importance in statistics are the moments of a distribution. Given a joint distribution $p$ over $N$ variables $\bm{X}= (X_1, \dots, X_N) \in \mathcal{X}$, and a set $S$ of integers that denote a subset $\bm{X}_S \coloneq \{X_i \mid i \in S\} \subseteq X$, the mixed moment of a set of variables $S$ is given by
\begin{align}
    \mathbb{E}(\prod_{i \in S} X_i) &= \sum_{x_1, \dots, x_N \in \mathcal{X}} p(X_1=x_1, \dots, X_N=x_N) \prod_{i \in S} X_i \\
    &\coloneq \langle S \rangle 
\end{align}
If we assume that there is no inherent ordering to the variables, then it is natural to decompose the moment into elemental contributions $e(t)$ of elements of the powerset mereology on $S$:
\begin{align}
    \langle S \rangle &= \sum_{T \in \mathcal{P}(S)} e(T)\\
    &= \sum_{T \subseteq S} e(T)
\end{align}
To find out what the contributions $e(t)$ are, we can invert this sum using a Möbius inversion. Note, however, that the quantities $X_i$ might be dimensionful quantities, so to make this sum well-defined, we interpret $\langle S\rangle$ as the product $ \langle \prod_{i \in S} X_i \rangle \prod_{i \in S\setminus S} \langle X_i \rangle$, and write the full Möbius inversion as
\begin{align}
    e(S) &= \sum_{T \subseteq S} \mu_{\mathcal{P}}(T, S) \langle T \rangle \prod_{S_i \in S \setminus T}  \langle X_i \rangle \\
    &= \sum_{T \subseteq S} (-1)^{|T|-|S|} \langle T \rangle \prod_{S_i \in S \setminus T}  \langle X_i \rangle 
    \intertext{which for the case of three variables $X_1, X_2, X_3$ yields}
    e(X_i) &= \langle X_i \rangle \\
    e(X_i, X_j) &= \langle X_i, X_j \rangle - \langle X_i \rangle \langle X_j \rangle\\
    e(X_1, X_2, X_3) &= \langle X_1, X_2, X_3 \rangle - \langle X_1, X_2 \rangle \langle X_3 \rangle - \langle X_1, X_3 \rangle \langle X_2 \rangle - \langle X_2, X_3 \rangle \langle X_1 \rangle + 2 \langle X_1 \rangle \langle X_2 \rangle \langle X_3 \rangle
\end{align}
These are exactly the mixed central moments (it can be readily verified that this construction generalises to all higher-order moments), so that central moments are the Möbius inverse of mixed moments with respect to the powerset mereology. 

What happens if we impose a different mereology on the moments? For example, one might decompose a moment into contributions from all possible partitions $\pi$ of the system:
\begin{align}
    \langle S \rangle &= \sum_{\pi \in \Pi(S)} \kappa(\pi) \label{eq:momentDecompPartition}
\end{align}
where now $\kappa(\pi)$ is the contribution of the partition $\pi$ to the mixed moment, and $\Pi(S)$ is the set of all partitions of $S$. To respect the dimensions of $X$ we define the moment of a partition to be the product of moments of the blocks, so we can invert \eqref{eq:momentDecompPartition} over the lattice of partitions ordered by refinement:
\begin{align}
    \kappa(S) &= \sum_{\pi \in \Pi(S)}  \mu_{\Pi}(\pi, S) \prod_{\pi_i \in \pi} \langle \pi_i \rangle \\
    &= \sum_{\pi \in \Pi(S)} (-1)^{|\pi|-1} (|\pi|-1)! \prod_{\pi_i \in \pi}\langle \pi_i \rangle
\end{align}
where we have written $S$ for the partition $\{S\}$. This happens to be the same as the central moments for up to three variables, but is different afterwards. The $\kappa$ are called the mixed cumulants, and famously offer an equivalent way to characterise the distribution $p$. 

Note that a decomposing a set of variables $S$ into subsets or partitions only makes sense if the variables commute. In the noncommutative case, it is more natural to decompose $S$ into so-called \emph{non-crossing} partitions. A partition is non-crossing if there is no chain of elements $A>B>C>D$ such that $A$ and $C$ are in the same block, $B$ and $D$ are in the same block, but $A$ and $D$ are in different blocks. For example, the partition $\{\{1, 2\}, \{3, 4\}\}$ is non-crossing, but $\{\{1, 3\}, \{2, 4\}\}$ is not. The Möbius function of the lattice of non-crossing partitions is known to be given by signed Catalan numbers, and can be used to define the noncommutative version of cumulants, called \emph{free cumulants} \cite{speicher1997free}. When sampling from stochastic processes, one might encounter path-valued variables. The statistics of such path-values samples can be summarised in a sequence of \textit{path signature moments}. The authors of \cite{bonnier2020signature} argue that the sequential nature of these path-valued variables makes it most natural to decompose them into \emph{ordered partitions}, which under their refinement order result in yet another mereology. Inverting the path signature moments over this ordered partition mereology yields the path signature cumulants, which are the natural generalisation of the cumulants to path-valued variables. These examples illustrate how different assumptions on the variables can be translated into different mereologies, which each lead to different notions of higher-order structure.

\bibliographystyle{unsrt}
\bibliography{refs}

\end{document}

%% file: booleanLatticeN2.tex
\begin{tikzpicture}[scale=1.5]

\coordinate[label={[xshift=1mm]right:$01$}] (11) at (0, 0);

\coordinate[label={[xshift=1mm]right:$0$}] (10) at (-1, -1);
\coordinate[label={[xshift=1mm]right:$1$}] (01) at (1, -1);
\coordinate[label={[xshift=1mm]right:$\emptyset$}] (00) at (0, -2);

\foreach \a/\b in {11/10, 11/01, 01/00, 10/00}
    \draw (\a) -- (\b);

\foreach \v in {11, 00, 01, 10}
    \draw[fill=black] (\v) circle (1pt);

\end{tikzpicture}

%% file: booleanLatticeN3.tex
\begin{tikzpicture}[scale=1.5]

\coordinate[label={[xshift=1mm]right:$012$}] (111) at (0, 0);

\coordinate[label={[xshift=1mm]right:$01$}] (110) at (-1, -1);
\coordinate[label={[xshift=1mm]right:$02$}] (101) at (0, -1);
\coordinate[label={[xshift=1mm]right:$12$}] (011) at (1, -1);

\coordinate[label={[xshift=1mm]right:$0$}] (100) at (-1, -2);
\coordinate[label={[xshift=1mm]right:$1$}] (010) at (0, -2);
\coordinate[label={[xshift=1mm]right:$2$}] (001) at (1, -2);

\coordinate[label={[xshift=1mm]right:$\emptyset$}, ] (000) at (0, -3);

\foreach \a/\b in {111/110, 111/101, 111/011, 110/100, 110/010, 101/100, 101/001, 011/010, 011/001, 100/000, 010/000, 001/000}
    \draw (\a) -- (\b);

\foreach \v in {000,001,010,011,100,101,110,111}
    \draw[fill=black] (\v) circle (1pt);

\end{tikzpicture}

%% file: booleanLatticeN4.tex
\begin{tikzpicture}[scale=1.5]

\coordinate[label={[xshift=1mm]right:$0123$}] (111) at (0, 0);

\coordinate[label={[xshift=1mm]right:$012$}] (110) at (-1, -1);
\coordinate[label={[xshift=1mm]right:$013$}] (101) at (0, -1);
\coordinate[label={[xshift=1mm]right:$023$}] (011) at (1, -1);

\coordinate[label={[xshift=1mm]right:$01$}] (100) at (-1, -2);
\coordinate[label={[xshift=1mm]right:$02$}] (010) at (0, -2);
\coordinate[label={[xshift=1mm]right:$03$}] (001) at (1, -2);

\coordinate[label={[xshift=1mm]right:$0$}, ] (000) at (0, -3);

\coordinate[label={[xshift=1mm]right:$123$}] (b111) at (3, -1);

\coordinate[label={[xshift=1mm]right:$12$}] (b110) at (2, -2);
\coordinate[label={[xshift=1mm]right:$13$}] (b101) at (3, -2);
\coordinate[label={[xshift=1mm]right:$23$}] (b011) at (4, -2);

\coordinate[label={[xshift=1mm]right:$1$}] (b100) at (2, -3);
\coordinate[label={[xshift=1mm]right:$2$}] (b010) at (3, -3);
\coordinate[label={[xshift=1mm]right:$3$}] (b001) at (4, -3);

\coordinate[label={[xshift=1mm]right:$\emptyset$}, ] (b000) at (3, -4);

\foreach \a/\b in {111/110, 111/101, 111/011, 110/100, 110/010, 101/100, 101/001, 011/010, 011/001, 100/000, 010/000, 001/000, b111/b110, b111/b101, b111/b011, b110/b100, b110/b010, b101/b100, b101/b001, b011/b010, b011/b001, b100/b000, b010/b000, b001/b000,000/b000,001/b001,010/b010,011/b011,100/b100,101/b101,110/b110,111/b111}
    \draw (\a) -- (\b);
\foreach \v in {000,001,010,011,100,101,110,111, b000,b001,b010,b011,b100,b101,b110,b111}
    \draw[fill=black] (\v) circle (1pt);

\end{tikzpicture}

%% file: partitionLatticeN2.tex
\begin{tikzpicture}
  \matrix (m) [matrix of math nodes, row sep=2em, column sep=2em] {
    12 \\
    1|2 \\
  };
  \path[-]
    (m-1-1) edge (m-2-1);
\end{tikzpicture}

%% file: partitionLatticeN3.tex
\begin{tikzpicture}
  \matrix (m) [matrix of math nodes, row sep=2em, column sep=.2em] {
    & 123 & \\
    13|2 & 1|23 & 12|3 \\
    & 1|2|3 & \\
  };
  \path[-]
    (m-1-2) edge (m-2-1)
            edge (m-2-2)
            edge (m-2-3)
    (m-2-1) edge (m-3-2)
    (m-2-2) edge (m-3-2)
    (m-2-3) edge (m-3-2);
\end{tikzpicture}

%% file: partitionLatticeN4.tex
\begin{tikzpicture}[scale=1.6]
  \node (m14) at (0,0) {1234};

  \node (m24) at (-3,-1) {$14|23$};
  \node (m25) at (-2,-1) {$1|234$};
  \node (m26) at (-1,-1) {$124|3$};
  \node (m27) at (0,-1) {$13|24$};
  \node (m28) at (1,-1) {$123|4$};
  \node (m29) at (2,-1) {$134|2$};
  \node (m210) at (3,-1) {$12|34$};

  \node (m34) at (-2.5,-2) {$1|23|4$};
  \node (m35) at (-1.5,-2) {$14|2|3$};
  \node (m36) at (-0.5,-2) {$1|24|3$};
  \node (m37) at (0.5,-2) {$13|2|4$};
  \node (m38) at (1.5,-2) {$12|3|4$};
  \node (m39) at (2.5,-2) {$1|2|34$};

  \node (m44) at (0,-3) {$1|2|3|4$};

  \path[-]
    (m14) edge (m24)
            edge (m25)
            edge (m26)
            edge (m27)
            edge (m28)
            edge (m29)
            edge (m210)
    (m24) edge (m34)
            edge (m35)
    (m25) edge (m34)
            edge (m36)
            edge (m39)
    (m26) edge (m35)
            edge (m36)
            edge (m38)
    (m27) edge (m36)
            edge (m37)
    (m28) edge (m34)
            edge (m37)
            edge (m38)
    (m29) edge (m35)
            edge (m36)
            edge (m38)
    (m210) edge (m38)
             edge (m39)
    (m34) edge (m44)
    (m35) edge (m44)
    (m36) edge (m44)
    (m37) edge (m44)
    (m38) edge (m44)
    (m39) edge (m44);
\end{tikzpicture}

%% file: corrFn.tex
\begin{equation}
    \langle X_1 X_2 X_3 X_4\rangle = 
    \begin{tikzpicture}[baseline=(current bounding box.mid),scale=0.4]
        \foreach \x/\y in {0/0, 1/0, 0/1, 1/1}
          \filldraw (\x, \y) circle (2pt);
        \foreach \x/\y/\i in {0/0/1, 1/0/2}
          \node at (\x, \y) [yshift=-5] {\tiny{$X_\i$}};
        \foreach \x/\y/\i in {0/1/3, 1/1/4}
          \node at (\x, \y) [yshift=4] {\tiny{$X_\i$}};
    \end{tikzpicture}
    +
    \begin{tikzpicture}[baseline=(current bounding box.mid),scale=0.4]
        \foreach \x/\y in {0/0, 1/0, 0/1, 1/1}
          \filldraw (\x, \y) circle (2pt);
        \draw (0,0) -- (1, 0);
    \end{tikzpicture}
    +
    \begin{tikzpicture}[baseline=(current bounding box.mid),scale=0.4]
        \foreach \x/\y in {0/0, 1/0, 0/1, 1/1}
          \filldraw (\x, \y) circle (2pt);
        \draw (0,0) -- (0,1);
    \end{tikzpicture}
    +
    \begin{tikzpicture}[baseline=(current bounding box.mid),scale=0.4]
        \foreach \x/\y in {0/0, 1/0, 0/1, 1/1}
          \filldraw (\x, \y) circle (2pt);
        \draw (1,0) -- (1,1);
    \end{tikzpicture}
    +
    \begin{tikzpicture}[baseline=(current bounding box.mid),scale=0.4]
        \foreach \x/\y in {0/0, 1/0, 0/1, 1/1}
          \filldraw (\x, \y) circle (2pt);
        \draw (0,1) -- (1,1);
    \end{tikzpicture}
    +
    \begin{tikzpicture}[baseline=(current bounding box.mid),scale=0.4]
        \foreach \x/\y in {0/0, 1/0, 0/1, 1/1}
          \filldraw (\x, \y) circle (2pt);
        \draw (0,0) -- (1,1);
    \end{tikzpicture}
    +
    \begin{tikzpicture}[baseline=(current bounding box.mid),scale=0.4]
        \foreach \x/\y in {0/0, 1/0, 0/1, 1/1}
          \filldraw (\x, \y) circle (2pt);
        \draw (1, 0) -- (0, 1);
    \end{tikzpicture}
    +
    \begin{tikzpicture}[baseline=(current bounding box.mid),scale=0.4]
        \foreach \x/\y in {0/0, 1/0, 0/1, 1/1}
          \filldraw (\x, \y) circle (2pt);
        \draw (0,0) -- (0,1);
        \draw (1,0) -- (1,1);
    \end{tikzpicture}
    +
    \begin{tikzpicture}[baseline=(current bounding box.mid),scale=0.4]
        \foreach \x/\y in {0/0, 1/0, 0/1, 1/1}
          \filldraw (\x, \y) circle (2pt);
          \draw (0,0) -- (1,0);
          \draw (0,1) -- (1,1);
    \end{tikzpicture}
    +
    \begin{tikzpicture}[baseline=(current bounding box.mid),scale=0.4]
      \foreach \x/\y in {0/0, 1/0, 0/1, 1/1}
        \filldraw (\x, \y) circle (2pt);
      \draw (0,0) -- (1,1);
      \draw (0, 1) to (0.35, 0.65);
      \draw (0.65, 0.35) to (1, 0);
    \end{tikzpicture}
    +
    \begin{tikzpicture}[baseline=(current bounding box.mid),scale=0.4]
        \foreach \x/\y in {0/0, 1/0, 0/1, 1/1}
          \filldraw (\x, \y) circle (2pt);
        \draw (0,0) -- (1,1);
        \draw (1,0) -- (0.5, 0.5);
      \end{tikzpicture}
    +
    \begin{tikzpicture}[baseline=(current bounding box.mid),scale=0.4]
        \foreach \x/\y in {0/0, 1/0, 0/1, 1/1}
          \filldraw (\x, \y) circle (2pt);
        \draw (0,1) -- (0.5, 0.5);
        \draw (0,0) -- (1,1);
    \end{tikzpicture}
    +
    \begin{tikzpicture}[baseline=(current bounding box.mid),scale=0.4]
        \foreach \x/\y in {0/0, 1/0, 0/1, 1/1}
          \filldraw (\x, \y) circle (2pt);
        \draw (0,1) -- (1, 0);
        \draw (0,0) -- (0.5, 0.5);
    \end{tikzpicture}
    +
    \begin{tikzpicture}[baseline=(current bounding box.mid),scale=0.4]
        \foreach \x/\y in {0/0, 1/0, 0/1, 1/1}
          \filldraw (\x, \y) circle (2pt);
        \draw (0,1) -- (1, 0);
        \draw (1, 1) -- (0.5, 0.5);
    \end{tikzpicture}
    +
    \begin{tikzpicture}[baseline=(current bounding box.mid),scale=0.4]
        \foreach \x/\y in {0/0, 1/0, 0/1, 1/1}
          \filldraw (\x, \y) circle (2pt);
        \draw (0,1) -- (1, 0);
        \draw (1, 1) -- (0, 0);
    \end{tikzpicture} \label{eq:corrFnDiags}
    \end{equation}